\documentclass{aa}
\usepackage{graphicx}
\usepackage{subfig}
\usepackage{txfonts}
\usepackage{amstext}
\usepackage{color}

\begin{document}

\newcommand{\va}{7~Vul }
\newcommand{\ve}{7~Vul}
\newcommand{\bu}{$\bullet$ }
\newcommand{\Mnom}{\hbox{$\mathcal{M}^{\mathrm N}_\odot$}}
\newcommand{\Rnom}{\hbox{$\mathcal{R}^{\mathrm N}_\odot$}}
\newcommand{\Lnom}{\hbox{$\mathcal{L}^{\mathrm N}_\odot$}}
\newcommand{\GMnom}{\hbox{$\mathcal{GM}^{\mathrm N}_\odot$}}
\newcommand{\spefo}{{\tt SPEFO} }
\newcommand{\spefoe}{{\tt SPEFO}}
\newcommand{\respefo}{{\tt reSPEFO} }
\newcommand{\respefoe}{{\tt reSPEFO}}

\newcommand{\phoebe}{{\tt PHOEBE} }
\newcommand{\phoebee}{{\tt PHOEBE}}
\newcommand{\fotel}{{\tt FOTEL} }
\newcommand{\fotele}{{\tt FOTEL}}
\newcommand{\korel}{{\tt KOREL} }
\newcommand{\korele}{{\tt KOREL}}
\newcommand{\binsyn}{{\tt BINSYN} }
\newcommand{\binsyne}{{\tt BINSYN}}
\newcommand{\pyt}{{\tt PYTERPOL} }
\newcommand{\pyte}{{\tt PYTERPOL}}

\newcommand{\tria}{\hbox{$\bigtriangleup$}}
\newcommand{\ubv}{\hbox{$U\!B{}V$}}
\newcommand{\ubvr}{\hbox{$U\!B{}V\!R$}}
\newcommand{\bv}{\hbox{$B\!-\!V$}}
\newcommand{\ub}{\hbox{$U\!-\!B$}}
\newcommand{\hp}{$H_{\rm p}$}

\newcommand{\p}{$\pm$}
\newcommand{\arcm}{$^\prime$}
\newcommand{\arcs}{$^{\prime\prime}$}
\newcommand{\m}{$^{\rm m}\!\!.$}
\newcommand{\D}{$^{\rm d}\!\!.$}
\newcommand{\F}{$^{\rm P}\!\!.$}
\newcommand{\kms}{km~s$^{-1}$ }
\newcommand{\ks}{km~s$^{-1}$}
\newcommand{\ms}{M$_{\odot}$}
\newcommand{\rs}{R$_{\odot}$}
\newcommand{\oc}{$O-C$}
\newcommand{\ha}{H$\alpha$ }
\newcommand{\hb}{H$\beta$ }
\newcommand{\hg}{H$\gamma$ }
\newcommand{\tef}{$T_{\rm eff}$ }
\newcommand{\teff}{$T_{\rm eff}$}
\newcommand{\lgg}{{\rm log}~$g$ }
\newcommand{\lgge}{{\rm log}~$g$}
\newcommand{\vsin}{$v$~sin~$i,$ }
\newcommand{\vsi}{$v$~sin~$i$}

\newcommand{\hae}{H$\alpha$}
\newcommand{\hbe}{H$\beta$}
\newcommand{\hge}{H$\gamma$}
\newcommand{\hde}{H$\delta$}
\newcommand{\he}{\ion{He}{i}~6678~\AA \ }
\newcommand{\hea}{\ion{He}{i}~6678~\AA}
\newcommand{\cii}{\ion{C}{ii} }
\newcommand{\Am}{\ANG~mm$^{-1}$ }
\newcommand{\Ame}{\ANG~mm$^{-1}$}

\newcommand{\hip}{$Hipparcos$}
\newcommand{\ond}{Ond\v{r}ejov}

\title{A new study of the spectroscopic binary 7~Vul with a Be star primary
\thanks{Based on new spectral and photometric observations from the following observatories: Dominion Astrophysical Observatory, Hvar, Ondřejov, Hipparcos, and ASAS3, ASAS-SN, KELT, and TESS services.
 } }
\author{P. Harmanec\inst{1}\and J. Lipt\'ak\inst{1}\and P.~Koubsk\'y\inst{2}\and
H.~Bo\v{z}i\'c\inst{3}\and J.~Labadie-Bartz\inst{4}\and
M.~\v{S}lechta\inst{2}\and S.~Yang\inst{5}\and A.~Harmanec\inst{6}
}
\institute{Astronomical Institute, Faculty of Mathematics and Physics,
           Charles University, V~Hole\v{s}ovi\v{c}k\'ach~2, CZ-180~00~Praha~8, \\
           Czech Republic, \email{hec@sirrah.troja.mff.cuni.cz}
      \and Astronomical Institute, Academy of Sciences, Fri\v{c}ova~298,
           CZ-251~65~Ond\v{r}ejov, Czech Republic
      \and Hvar Observatory, Faculty of Geodesy, Zagreb University,
           Ka\v ci\'ceva~26, HR-10000 Zagreb, Croatia
\and Instituto de Astronomia, Geof\'\i sica e Ciencias Atmosf\'ericas,
Universidade de S\~ao Paulo, Rua do Mat\~ao 1226, Cidade Universit\'aria,
05508-900 S\~ao Paulo, SP, Brazil
\and Physics \& Astronomy Department, University of Victoria,
           PO Box 3055 STN CSC, Victoria, BC, V8W 3P6, Canada
      \and Undergraduate student,
           Faculty of Mathematics and Physics,
           Charles University, V~Hole\v{s}ovi\v{c}k\'ach~2, CZ-180~00~Praha~8,
           Czech Republic,
}
\date{Received \today}

\abstract {We confirmed the binary nature of the Be star 7~Vul, derived
a~more accurate spectroscopic orbit with an orbital period of
$69\fd4212\pm0\fd0034,$ and improved the knowledge
of the basic physical elements of the system. Analyzing available photometry
and the strength of the \ha emission, we also document the long-term spectral
variations of the Be primary. In addition, we confirmed rapid light changes with
a~period of 0\fd5592, which is comparable to the expected rotational period
of the Be primary, but note that its amplitude and possibly its period
vary with time. We were able to disentangle only
the \ion{He}{i}~6678~\AA\ line of the secondary, which could support
our tentative conclusion that the secondary appears to be a hot subdwarf.
A search for this object in high-dispersion far-UV spectra
could provide confirmation.
Probable masses of the binary components are ($6\pm1$)~\Mnom \ and
($0.6\pm0.1$)~\Mnom. If the presence of a hot subdwarf is firmly confirmed,
7~Vul might be identified as a rare object with a B4-B5 primary; all Be + hot subdwarf systems found so far contain B0-B3 primaries.}


\maketitle

\section{Introduction}
The B5 star \va (HR~7409, HD~183537, HIP~95818, Boss~ 4981; $V$=6\m3-6\m4 var.) was recently discovered to be a Be star and a spectroscopic binary with a 69~d orbital period and a low-mass companion, not readily seen in the spectra \citep{vennes}. A modest history of the previous investigations of this object was summarised in that study and here we only
add a few brief comments. \citet{hall70} studied cluster Collinder~399 and concluded that \va is not a member. When Hipparcos parallaxes were released, \citet{skiff98} and \citet{baum98}  independently found that the cluster does not exist at all.

Since the publication of \citet{vennes}, new spectra of the star have been accumulated in Ond\v{r}ejov (OND), the Dominion Astrophysical Observatory (DAO), and in the BeSS database. It was therefore deemed useful to carry out a new study of the object using modern methods of data analysis. Moreover, we also secured \ubv\ and \ubvr\ photometry of the star at Hvar and collected, homogenised, and analysed photometric observations from several available sources.

\section{Observational data and their reduction}
Throughout this paper, we specify all times of observations using
reduced heliocentric Julian dates

\smallskip
\centerline{RJD = HJD -2400000.0\,}
\smallskip\noindent
to avoid a possible 0.5~d confusion caused by the use of modified Julian dates (MJD). We also use the nominal values of the solar units and numerical
constants recommended by \citet{units2016}.

\subsection{Spectroscopy}
Our observational material consists of 81 \ond\ CCD spectra
secured in 2003--2011, 5 DAO spectra, 16 \ond\ CCD spectra secured in June-August 2019,
and 14 spectra secured by experienced amateurs and provided via BeSS database \citep{neiner2011}. See
\url{http://basebe.obspm.fr/basebe/} for details
on the instruments and observers.
A journal of these observations with basic information is provided in Table~\ref{jourv}.
\begin{table*}
\caption{Journal of spectroscopic observations of \ve.}
\begin{center}
\begin{tabular}{ccrccc}
\hline\hline\noalign{\smallskip}
Spg. & RJD          & No.        & Wavelength & $S/N$  & Spectral\\
     & range        &            & range      &range  & resolution\\
\hline\noalign{\smallskip}
A & 54244.45-55906.23 & 81 & 6258-6770 & 63-485 &12700 \\
B & 54276.82-54443.62 &  5 & 6155-6763 &155-266 &17200 \\
C & 58640.37-58714.36 & 16 & 6263-6735 &124-551 &12700 \\
\hline\noalign{\smallskip}
\multicolumn{6}{c}{BeSS amateur spectra}\\
\hline\noalign{\smallskip}
D & 56480.48-57579.53 &  4 & 6500-6610 & 37-72 &13400 \\
E &    56508.40       &  1 & 6490-6640 &   52  &13700 \\
F & 57205.38-57970.40 &  2 & 4190-7310 & 77-102&11000 \\
G & 57207.57-58258.49 &  2 & 6530-6690 & 80-145&15000 \\
H & 57676.56-58729.63 &  2 & 6500-6650 & 29-64 &10100 and 12700 \\
I & 58310.44-58663.48 &  3 & 4200-7360 & 73-117&11000 \\
\hline\noalign{\smallskip}
\end{tabular}
\tablefoot{Column ``Spg." identifies individual spectrographs and detectors used:
 A: OND 2.00~m reflector, coud\'e grating spg., CCD SITe5 2030 $\times$ 800 pixel detector;
 B: DAO 1.22~m reflector, McKellar spectrograph, 4K$\times$2K SITe CCD detector;
 C: OND 2.00~m reflector, coud\'e grating spg., CCD Pylon Excelon 2048 $\times$ 512 pixel detector;
 D: Dijon C8 LHIRES3 ATIK314L+, observer A.~Favaro;
 E: Haute Provence C9 LHIRES3 2400 ATIK460EX, observer V.~Desnoux;
 F: Tourbiere RC400 Astrosib-Eshel-ATIK460EX, observer O.~Garde;
 G: Verny C11 LHIRES3\# 194-2400t35-QSI516S, observer F.~Houpert;
 H: Manhattan lx200 12" LHIRES 2400 35u ATIK460EX, obsever K.~Graham;
 I: Revel eShel ATIK460, observer O.~Thizy.
 }
\end{center}
\label{jourv}
\end{table*}

An initial reduction of all spectra to 1D frames was carried out at
the respective observatories. To their normalisation and radial-velocity (RV)
measurements, we used the new Java program \respefoe\  written by one of
the authors of the present study (AH). The program allows RV measurements
via a comparison of direct and flipped images of line profiles. This is
especially convenient in situations where one aims to measure RVs of complicated \ha profiles, where emission is also present. Moreover, this also gives the chance to see and avoid disturbances from - often quite strong -
telluric lines. Details of the program are outlined in Appendix~\ref{apc},
while some additional comments on the spectral reductions are in
Appendix~\ref{apb}.

\subsection{Photometry}
We found only four series of earlier \ubv\ observations, published regrettably as the mean values only, without accurate dates.
\citet{craw71} published the mean of three all-sky \ubv\ observations secured sometime between RJD~37700 and 40900.
\citet{hall70} give mean values for an approximate epoch RJD~40500.
\cite{chamb77} also published mean values from 17 nights of observations for an approximate epoch RJD~42700.
\citet{yama77} obtained 100 all-sky observations in 18 nights between
Oct~22 and Dec~26, 1976 (RJD $\sim 43100$).

Table 2 presents a journal of photometric observations with known times  of observations, which were either obtained in the Johnson \ubv\ system or could be transformed to it. Details of the observations,
data reduction, and homogenisation are in Appendix~\ref{apa}.

\begin{table*}
\caption{Journal of photometric observations of \ve.}
\begin{center}
\begin{tabular}{rcrcrc}
\hline\hline\noalign{\smallskip}
Station & RJD          & No.        &Photometric& Comparison&Ref.\\
        & range        &            & system    &  star      &    \\
\hline\noalign{\smallskip}
 10&(37700-40900)     &  3 & \ubv & all-sky     & 2\\
 42& $\sim40500$      &  3 & \ubv & 5 Vul       & 1\\
 13&41497.87-41938.71 & 38 & DAO  & V395 Vul    & 3\\
112& $\sim42700$      &  ? & \ubv & all-sky     & 4\\
113&(43074-43139)     &100 & \ubv & all-sky     & 5\\
 61&47959.77-48796.88 & 99 &\hp   & all-sky     & 6\\
 93&52724.90-54755.52 &    & $V$  & all-sky     & 7\\
115&54257.77-56457.88 &3614&KELT $R$& all-sky   &10\\
  1&54273.41-55726.52 & 92 & \ubv &13 Vul, 9 Vul& 8\\
114&57062.17-58427.71 &574 & $V$  & all-sky     & 9\\
114&58220.88-58809.55 &    & $g$  & all-sky     & 9\\
  1&58665.43-58691.46 & 32 & \ubvr&13 Vul, 9 Vul& 8\\
110&58683.37-58710.18 &1237&TESS 5860-11260~\AA \ band&all-sky&11\\
\hline\noalign{\smallskip}
\end{tabular}
\tablefoot{Individual observing stations are distinguished by
the running numbers they have in the Prague / Zagreb photometric
archives --- see column "Station".
01:~Hvar 0.65~m, Cassegrain reflector, EMI9789QB tube;
10:~Kitt Peak National Observatory 0.40~m reflector, cooled 1P21 tube;
13:~DAO 0.40~m reflector, DAO photometry;
42:~Dyer observatory 0.61~m Cassegrain reflector, 1P21 tube;
61:~Hipparcos all-sky $H_{\rm p}$ photometry transformed to Johnson $V$;
93:~ASAS data archive \citep{pojm2002};
110:~TESS satellite \citep{Ricker2016};
112:~Kutztown State College Observatory 0.46 m Cassegrain reflector, EMI~6256A tube;
113:~Mitaka 0.30~m reflector, Toshiba 1P21 tube;
114:~ASAS-SN photometric network;
115:~Winer Observatory KELT 0.042~m wide-field survey telescope with
     a~CCD camera \citep{Pepper2007}.\\
 Column "Ref.":
1:~\citet{craw71}
	2:~\citet{hall70};
	3:~\citet{hill76};
	4:~\citet{chamb77};
	5:~\citet{yama77};
	6:~\citet{esa97};
	7:~\cite{pojm2002};
	8:~this paper;
	9:~\citet{asas2014,asas2017,asas2019};
	10:~\citet{bartz2017};
	11:~this paper.
}
\end{center}
\label{jouphot}
\end{table*}

\section{Long-term changes}
\begin{figure}
\centering
\resizebox{\hsize}{!}{\includegraphics{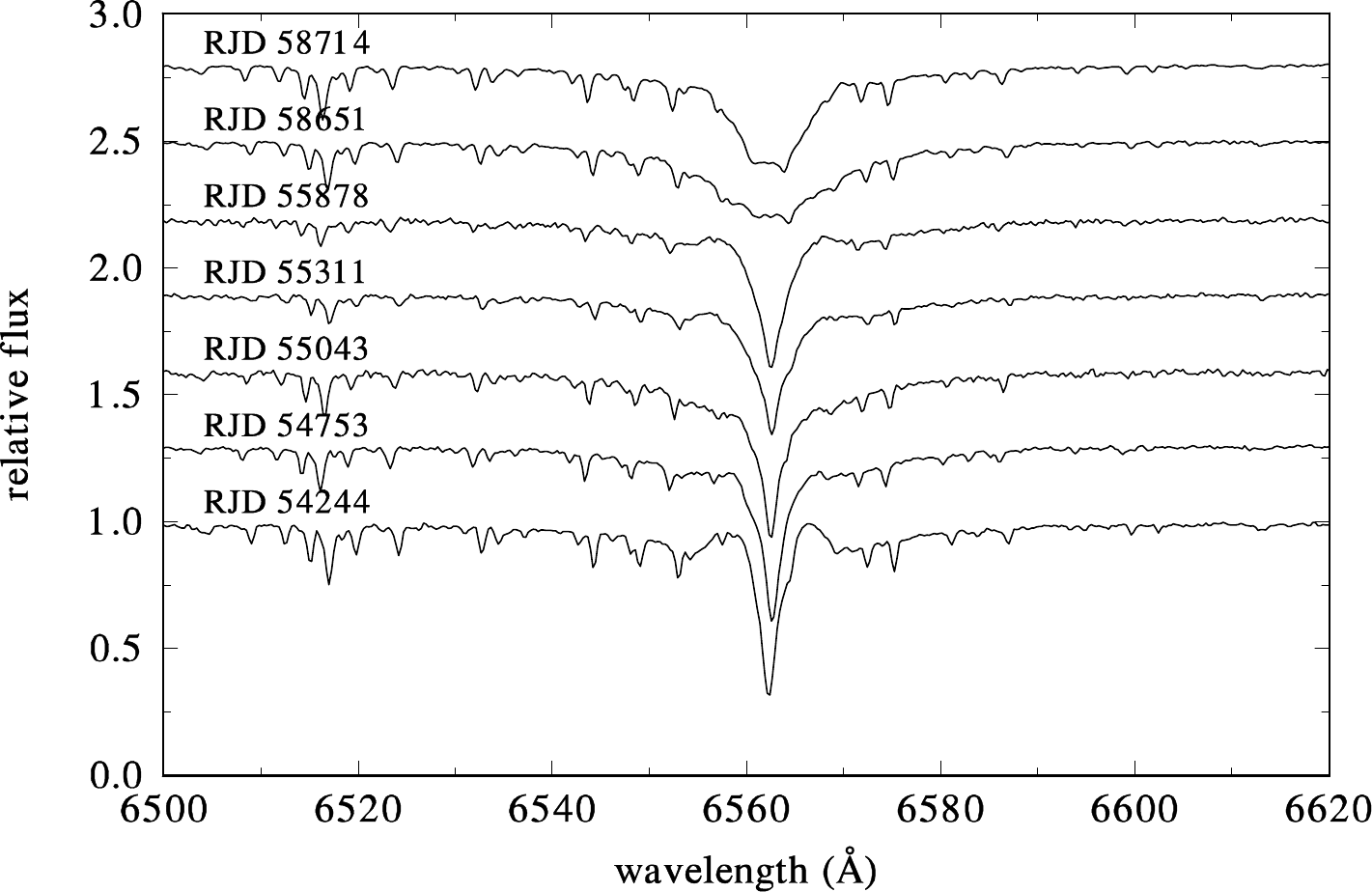}}
\caption{Selection of Ond\v{r}ejov \ha profiles, illustrating the gradual disappearance of the double emission and later also of the shell core.}\label{h3prof}
\end{figure}

\begin{figure}
\centering
\resizebox{\hsize}{!}{\includegraphics{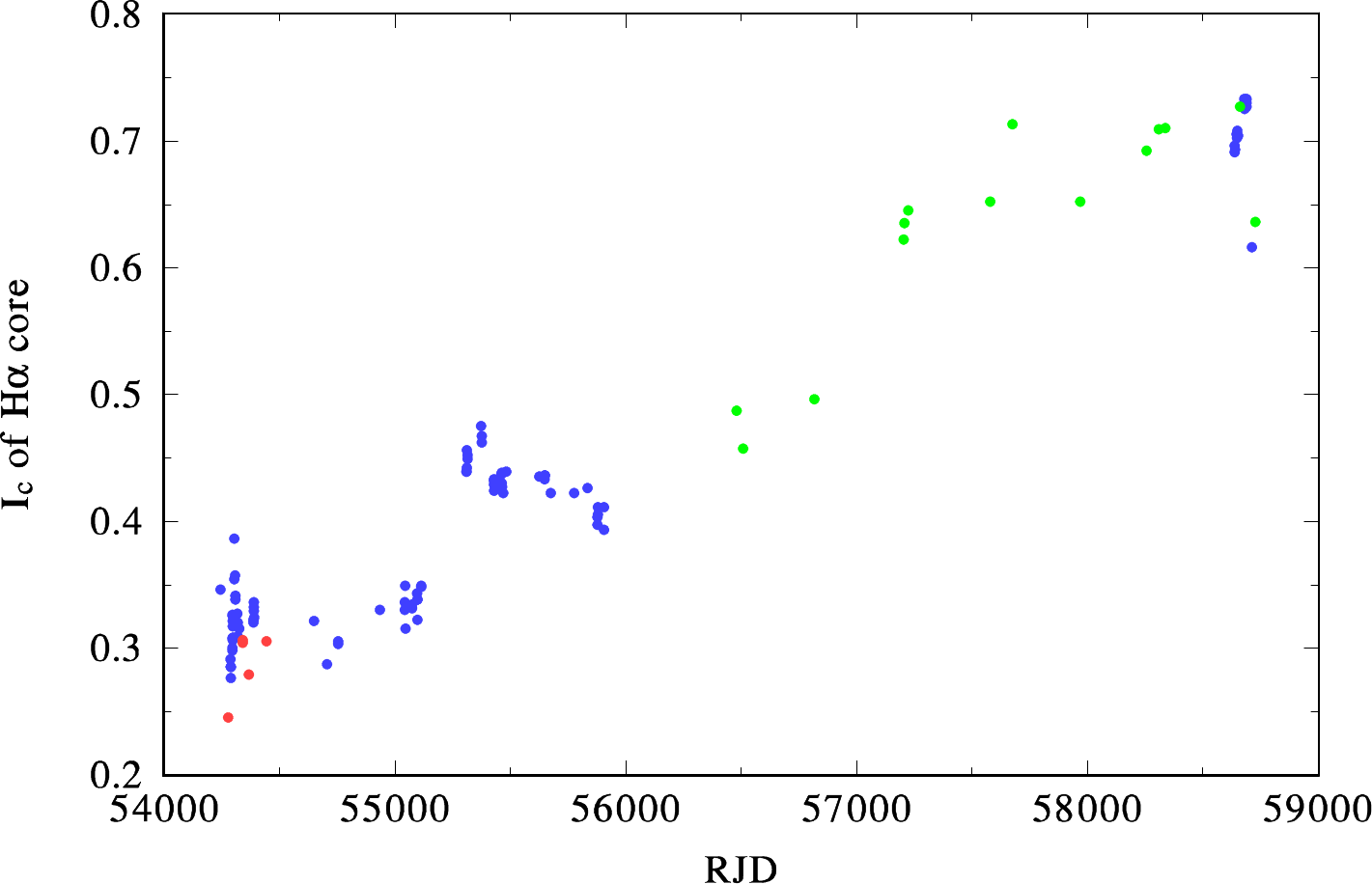}}
\resizebox{\hsize}{!}{\includegraphics{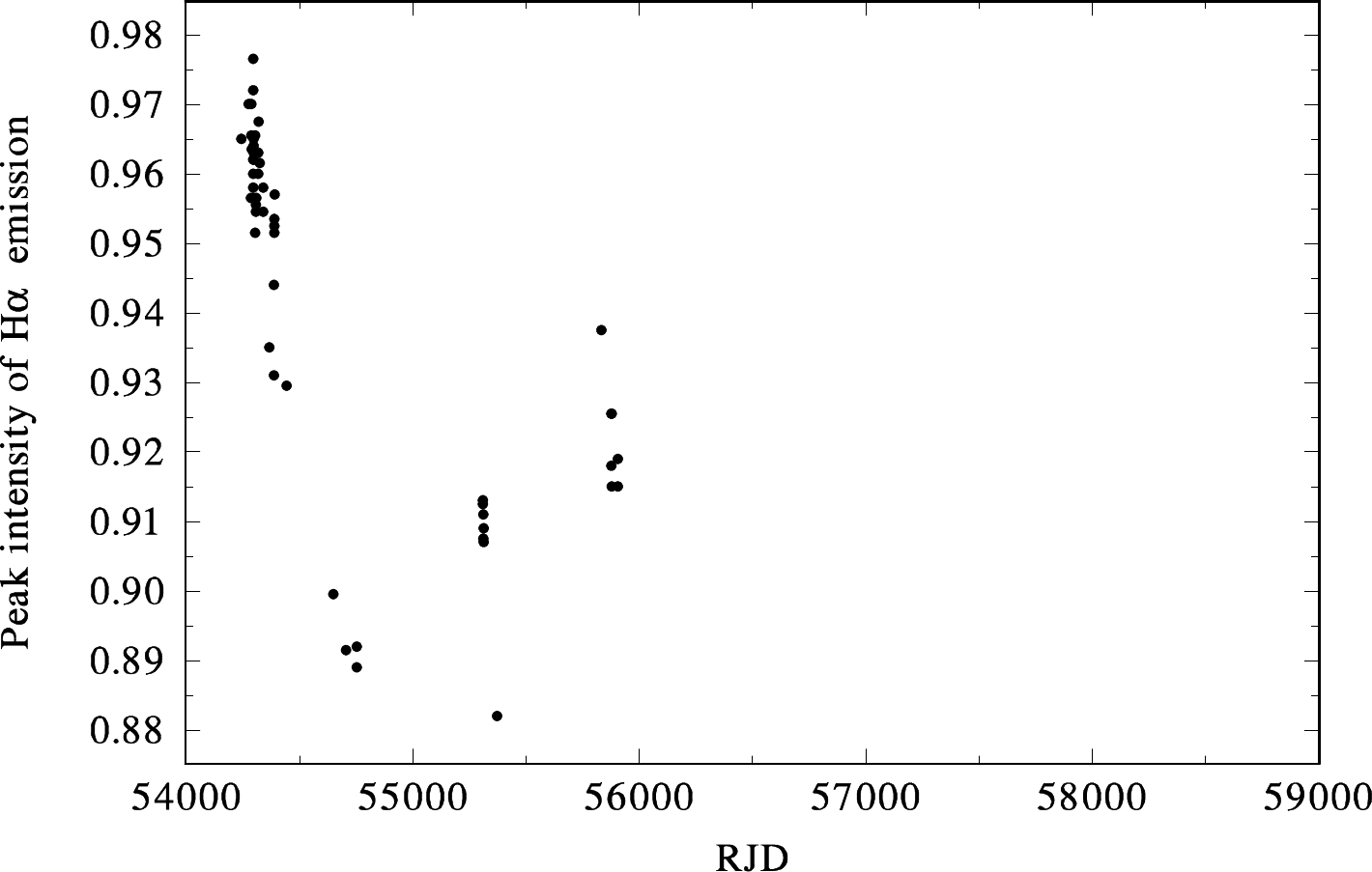}}
\caption{Top: Evolution of the central intensity of \ha absorption core over the time interval covered by available data. Individual datasets are distinguished as follows:
Red: DAO spectra; blue: Ond\v{r}ejov spectra, green: amateur spectra from the BeSS database.
Bottom: Time evolution of the peak intensity ($V+R$)/2 of the \ha emission over the same interval of time.}\label{ictime}
\end{figure}

In spite of the relatively short record of spectral and light variations, \va seems to be quite active. In 2007, when it was discovered as a Be star, it had a double \ha emission rising slightly below the continuum level. The emission gradually disappeared during 2008, and was absent in 2009. Weaker emission temporarily re-appeared in April 2010. Another weak emission episode was recorded by the end of 2011. Throughout that time, a shell core
of \ha was clearly visible. The star was completely
without detectable circumstellar matter, appearing as a normal rapidly rotating Be star in the first half of 2019. However, in the spectrum from August 2019, a \ha shell line is again seen. A representative selection of \ha line profiles is shown in Fig.~\ref{h3prof}.
We measured the central absorption peak of the \ha profile in all spectra at our disposal. Whenever the double emission was seen, we also measured the $V$ and $R$ peak intensities. Figure~\ref{ictime} shows the time evolution of the central intensity
and of the peak emission $(I_{\rm V}+I_{\rm R})/2$. Good agreement
can be seen between the values from different spectrographs. In addition to a secular weakening of the \ha absorption core, clear variations can be seen on a shorter timescale.
\begin{figure}
\centering
\resizebox{\hsize}{!}{\includegraphics{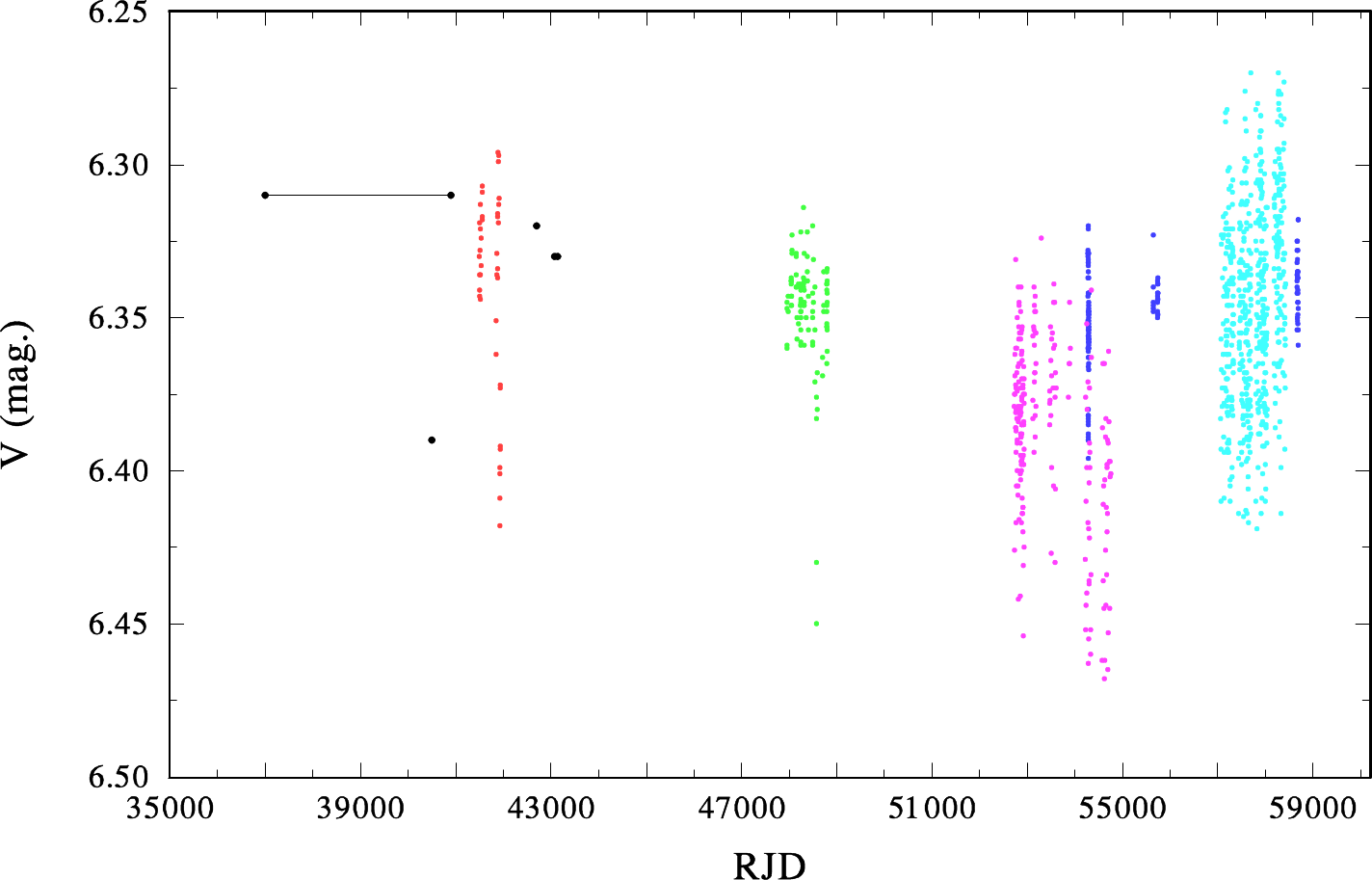}}
\resizebox{\hsize}{!}{\includegraphics{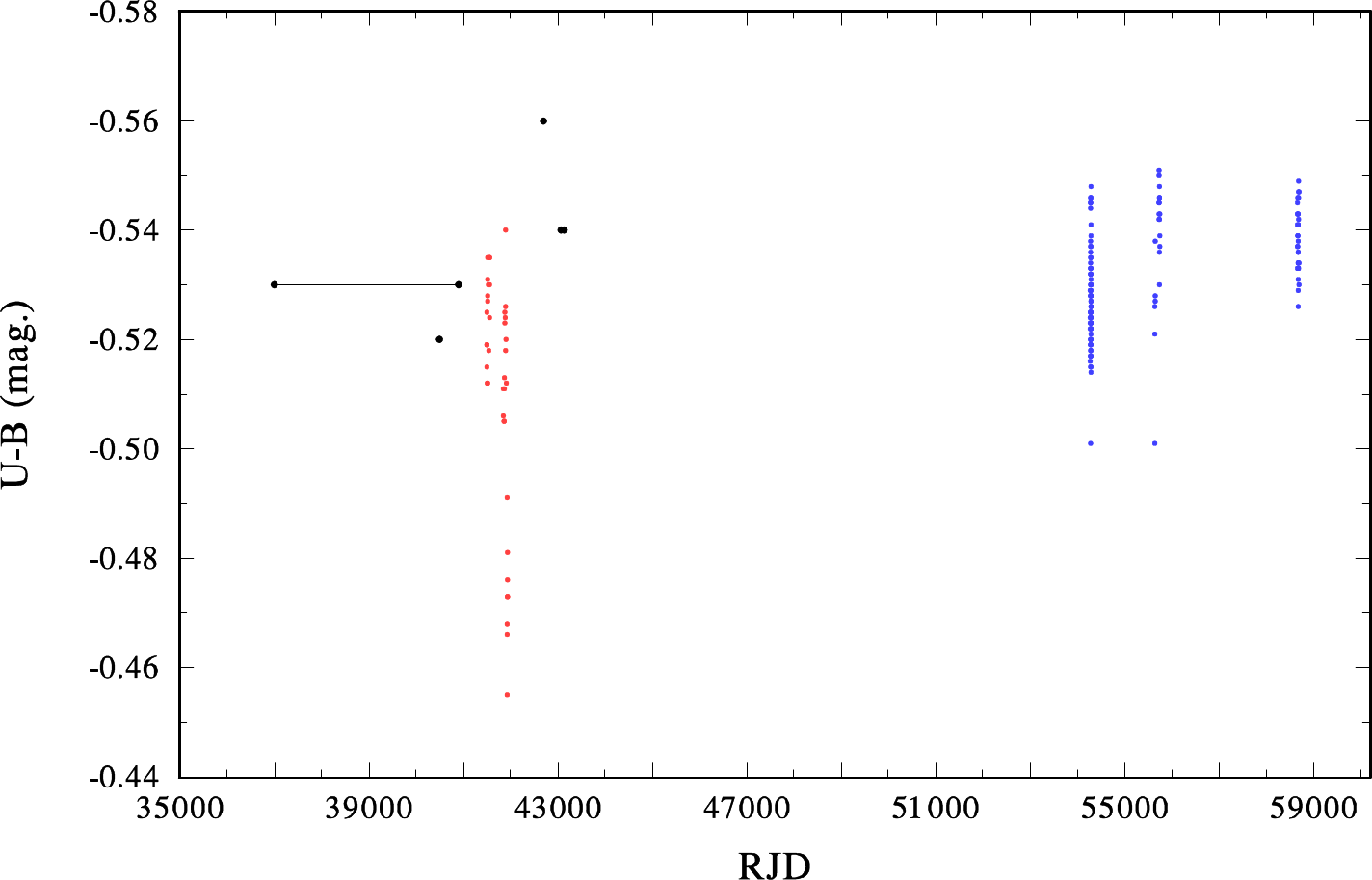}}
\caption{Top: Variation of the $V$ magnitude over the time interval covered by available data. Bottom: Variation of the \ub\ colour over the same time interval. In both panels, data from individual observing stations listed in Table~\ref{jouphot} are shown by coloured circles as follows:
01: blue; 10, 42, 112, and 113 (with uncertain dates of observations): black;
13: red; 61: green, 93: magenta, 114: cyan.}\label{ubvtime}
\end{figure}
The long-term variations in brightness and colour (Fig.~\ref{ubvtime})
are not very pronounced. We suspect some brightness decrease in the interval RJD~$\sim 53000-55000$, when the \ha emission was present in the spectra. If real, this would indicate an inverse correlation between the brightness and emission-line strength as defined by \citet{hec83}. This type of correlation
is observed for objects seen more or less equator-on. The increase of
the emission strength is accompanied by a light decrease. In the \ub \ versus
\bv \ diagram, the object moves along the main sequence towards a later
spectral subclass.

\begin{figure}
\centering
\resizebox{\hsize}{!}{\includegraphics{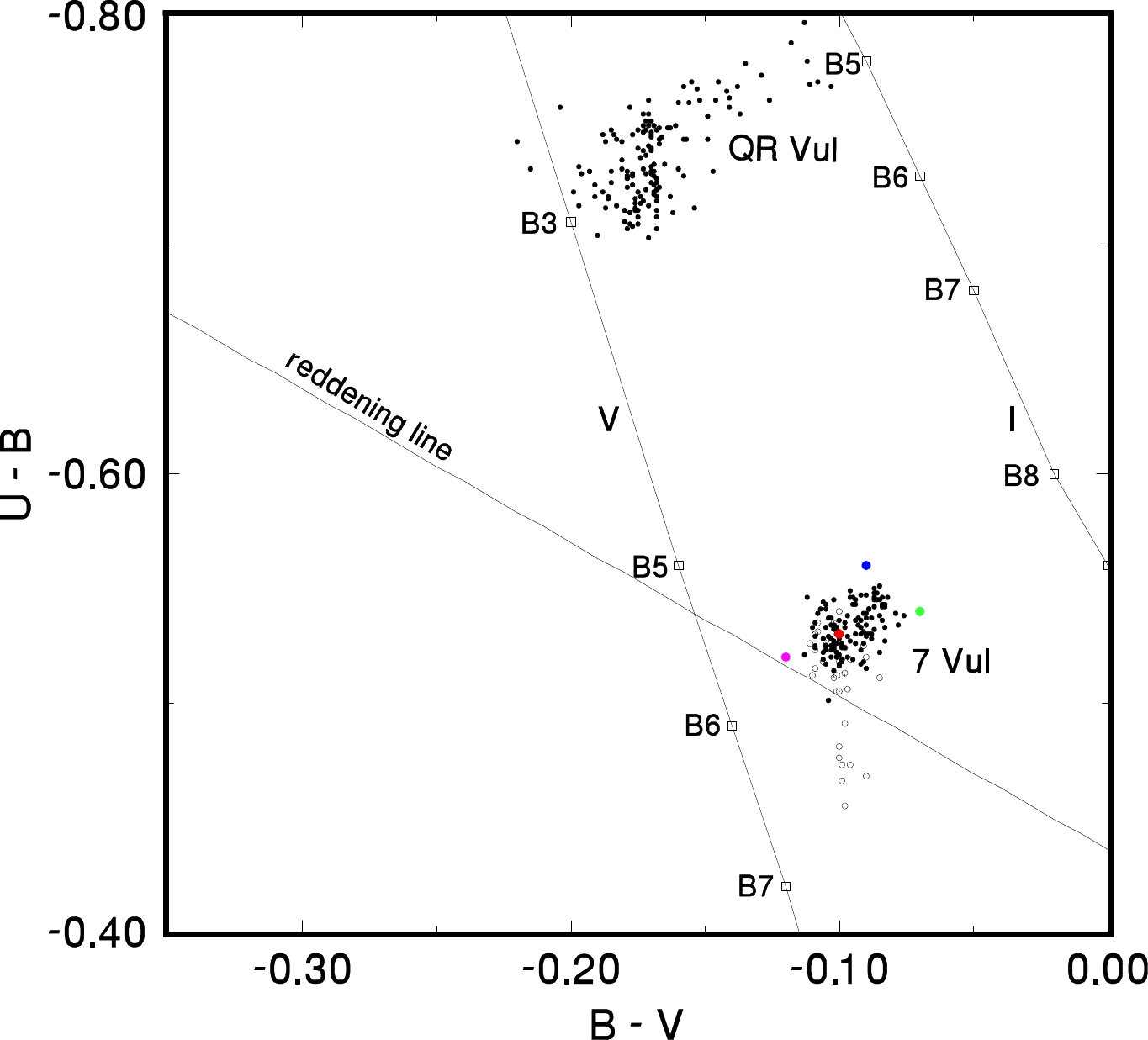}}
\caption{Variation of \va in the \ub\ vs. \bv\ diagram. Also shown for
comparison are the variations of another Be star observed at Hvar:
QR~Vul. Data from individual observing stations listed in Table~\ref{jouphot} are shown by coloured circles as follows:
01: black; 10: red; 13: open black circles; 42: magenta; 112: blue; 113: green. We note that the peculiar behavior of observations from station 13 might be due to variability of the comparison star V395~Vul, and not variability
of \va itself.}\label{ubbv}
\end{figure}

On the other hand,
the colour--colour variations shown in Fig.~\ref{ubbv} seem to indicate
a positive correlation for stars observed more pole-on, during which
the emission-strength increase is accompanied by a light brightening and
a shift of dereddened colours from the main sequence to giants or
supergiants of the same spectral class. This simple geometrical interpretation
was confirmed by the models published by \citet{sigut2013}.

\section{Orbital variations}
\subsection{An improved ephemeris}
Given the presence of secular spectral variations described in the previous section, it is not easy to apply some of the more sophisticated methods of RV measurement. This is why we opted for classical RV measurements using \respefoe. It is also important to mention that the vast majority of our data cover only the red part of the spectrum. OND and DAO spectra contain only four stronger lines:  \hae, \ion{Si}{ii}~6347 and 6371~\AA, and \ion{He}{i}~6678~\AA. The majority of the BeSS spectra cover only the \ha region; some also cover \ion{He}{i}~6678~\AA.
Only five BeSS spectra are echelle spectra, covering also blue parts of the spectrum. Since the RV zero-point can only be under control in the red region containing telluric lines, we preferred not to use RVs from the blue spectra. We are thus left with \hae, the only strong line available. For the spectra from epochs when the circumstellar matter was present, we were able to measure a sharp absorption core of \hae, while for the epochs without perceptible circumstellar matter, we could only measure the lower parts of the broader \ha absorption profile, with a lower accuracy than that from the sharp cores.
These RVs are listed in Table~\ref{newrv} in Appendix~\ref{apb}.
\begin{figure}
\centering
\resizebox{\hsize}{!}{\includegraphics{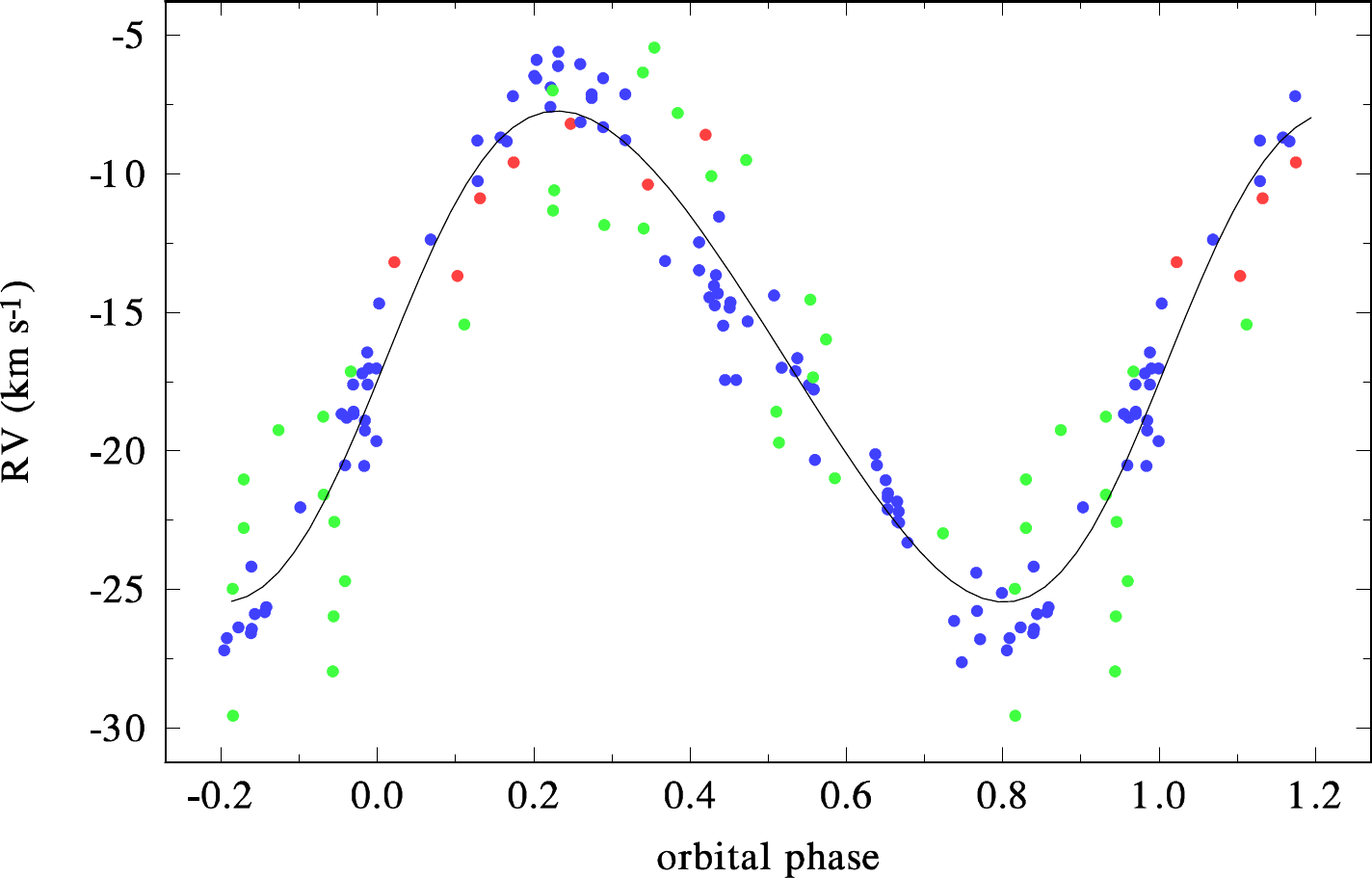}}
\resizebox{\hsize}{!}{\includegraphics{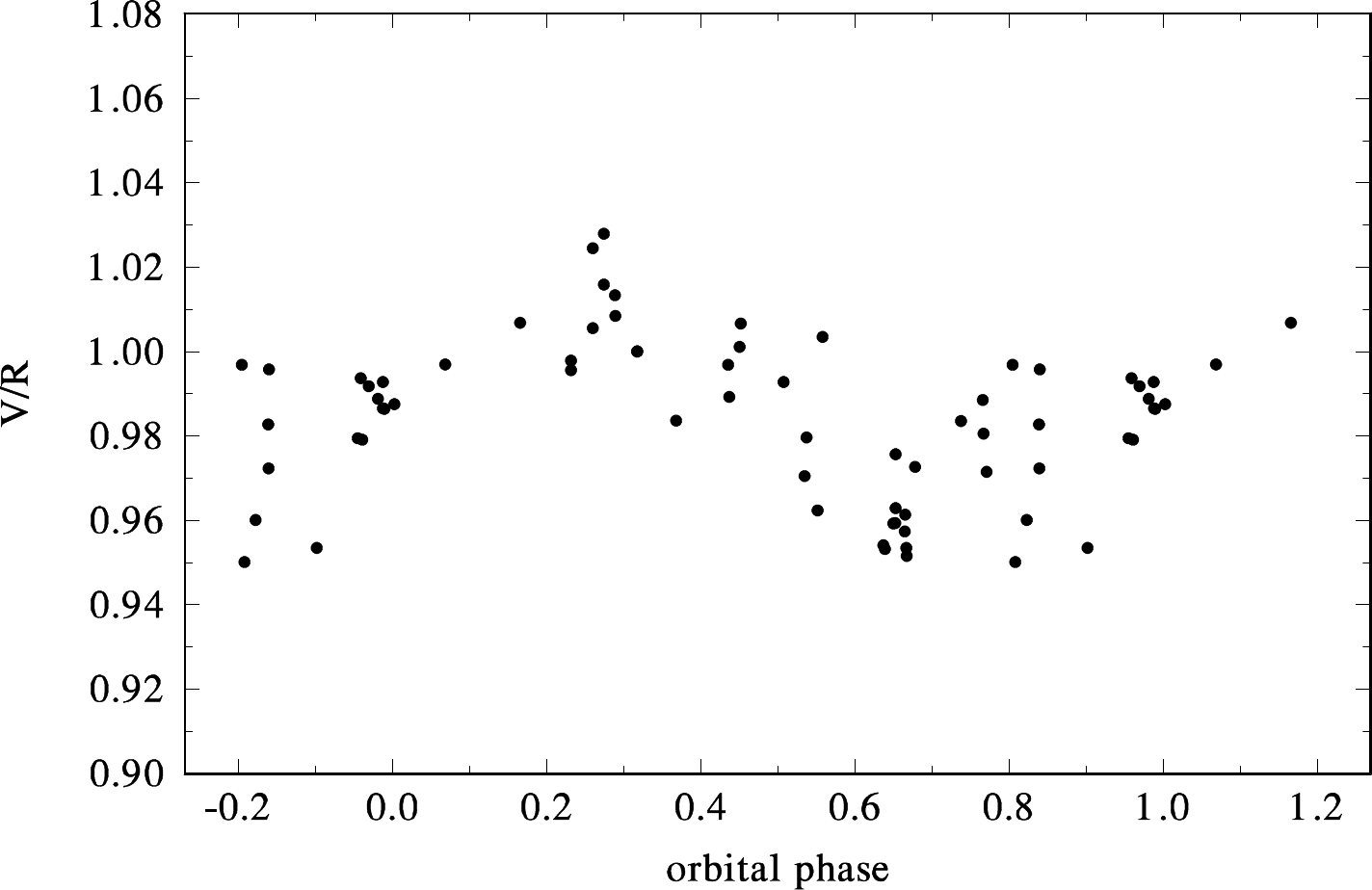}}
\caption{Top: RV curve based on all 124 RVs of the \ha absorption and old DAO
mean RVs \citep{plaskett}. We highlight the old DAO RVs (red;
decreased in RV for the difference between the systemic velocities of
these, and new spectra), new DAO and Ond\v{r}ejov RVs based on a sharp line
core from epochs where significant circumstellar matter is present (blue),
and Ond\v{r}ejov and BeSS spectra, based on RVs of broad \ha profiles without
perceptible emission (green). Our solution of Table~\ref{orbit} is shown
by a solid line.
Bottom: Phase-locked variations of the $V/R$ ratio of the double-peaked
\ha emission. For both plots, ephemeris~(1) was used.}\label{rvcurve}
\end{figure}

In addition to our RVs, we also used seven old DAO RVs secured between 1919 and 1925 and published as mean values of 5 to 12 lines \citep{plaskett}.
In cases where some of the plates were measured several times,
we adopted the mean RV of these measurements. For convenience, we provide these
RVs together with their RJDs in Table~\ref{oldrv} of Appendix~\ref{apb}.
\citet{plaskett} concluded that the star has a constant velocity but we verified that all their spectra were obtained in phases close to the RV maximum of the 69.4~d orbit.
Using all \ha absorption RVs and the old mean DAO RVs \citep{plaskett}, we calculated several trial orbital solutions. To this aim, we used the program \fotel \citep{fotel}. Our new orbital elements are compared to those by \citet{vennes} in Table~\ref{orbit} and the corresponding phase plot is in Fig.~\ref{rvcurve}. Our solution defines a new linear ephemeris
\begin{equation}
T_{\rm periastr.passage}={\rm RJD\,} 55986.8 + 69\fd4212\times E
,\end{equation}

\noindent to be used throughout this paper.

\begin{table}
\caption[]{The orbital solutions for the primary component of \ve.}
\label{orbit}
\begin{center}
\begin{tabular}{rcccccl}
\hline\hline\noalign{\smallskip}
Element            &  \citet{vennes}        & This paper      \\
\noalign{\smallskip}\hline\noalign{\smallskip}
$P$ (d)            &69.30\p0.07             & 69.4212\p0.0034 \\
$T_{\rm periastr.}$&54248.1\p2.7            &55986.8\p2.9     \\
$T_{\rm super.c .}$&54219.4                 &55953.5          \\
$e$                & 0.161\p0.035           & 0.113\p0.032    \\
$\omega$ ($^\circ$)& 247\p16                &264\p16          \\
$K_1$ (\ks)        &8.9\p0.4                & 8.86\p0.62      \\
$\gamma_1$ (\ks)   &    --                  &   $-$49.31\p0.89\\
$\gamma_2$ (\ks)   &  $-$14.8\p0.2          &   $-$16.51\p0.15\\
$\gamma_3$ (\ks)   &    --                  &   $-$16.24\p0.64\\
rms (\ks)         & not given               & 2.13            \\
No. of RVs        &   34                    & 123             \\
\noalign{\smallskip}\hline\noalign{\smallskip}
\end{tabular}
\tablefoot{All epochs are in RJD;
rms is the rms per 1 observation of unit weight. The systemic velocities
$\gamma$ of individual spectrographs are identified as follows: 1...old DAO
mean velocities; 2...sharp \ha cores during the emission phases; 3...broader \ha
absorption outside emission phases.}
\end{center}
\end{table}

We note that the rms errors of individual data sets are 2.17~\kms for the old DAO RVs, 1.40~\kms for the RVs of sharp-lined profiles, and 3.44~\kms for the RVs from broad \ha profiles. It is very encouraging that in spite of somewhat larger scatter,
the RVs from more or less photospheric profiles follow the same orbit and have the same systemic velocity within the error limits. This is strong evidence that the periodic changes are indeed caused by the orbital motion. However, a~word of warning is appropriate here. \citet{sterne41} and \citet{hec2003} pointed out
that a disturbance of line profiles that is symmetric with respect to line joining
the components can lead to spurious eccentricities with $\omega$ of either $90^\circ$
or $270^\circ$. This could be the case for \ve. We note that the orbital elements
of \va are quite similar to those of another Be star, V744 Her = 88~Her
\citep{hec74, doazan82a}. An eccentric orbit with $\omega\sim270^\circ$
was also found. A spurious eccentricity was detected for V832~Cyg = 59~Cyg
\citep{zarf21}. Another good example is the Be binary BR~CMi \citep{zarf30}, which is also
an ellipsoidal variable. A formal solution of its RV curve leads to an eccentric orbit
with $\omega\sim90^\circ$ but the light curve confirms a circular orbit.
Therefore, the possibility that the true orbit of \va is essentially circular should be kept in mind.
Another fact worth mentioning is that the old DAO RVs are systematically much more negative that the present-day RVs. Long-term RV changes with amplitudes larger than the amplitudes due to orbital motion are known for a number of Be stars, such as for example $\zeta$~Tau \citep{delplace70} or $\gamma$~Cas \citep{zarf20}. This could also be the case for \ve.
 Regrettably, no \ha profiles from the times of old DAO observations are available.
 We also mention that in some dynamical phases of the envelope evolution,
the RVs of shell lines can exhibit positive or negative RV shifts with
respect to their orbital RV, which could also distort the orbital RV curve
based on RVs from the shell lines.

For spectra with the presence of double \ha emission, we also derived the $V/R$ ratio of the emission peaks. As is seen in the bottom panel of Fig.~\ref{rvcurve}, the $V/R$ changes also vary with the orbital period and in phase with the RV changes. This can easily be understood by the fact that the `nose' of the Roche lobe around the primary is filled by a larger amount of emitting circumstellar material. Relatively
large scatter of the data around a mean trend is clearly due to the fact that the emission was relatively weak throughout its presence. Phase-locked $V/R$ changes constitute another proof of the binary nature of \ve. Similar variations were also found for V744~Her \citep{doazan82b}, V839~Her = 4~Her \citep{zarf6} and later for a number of other binary
Be stars \citep[cf., e.g.][]{rivi2013}.

\begin{figure}
\centering
\resizebox{\hsize}{!}{\includegraphics{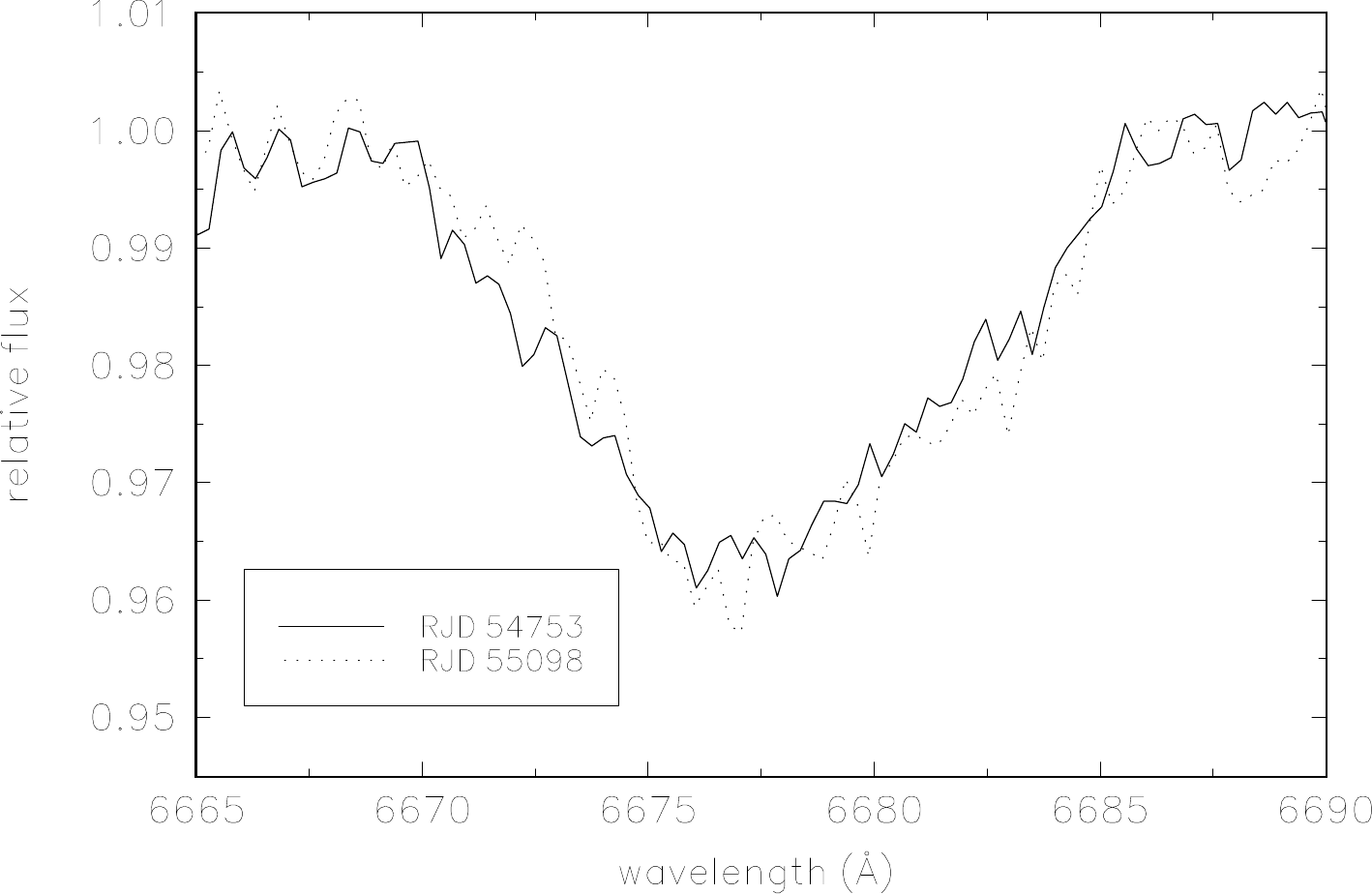}}
\resizebox{\hsize}{!}{\includegraphics{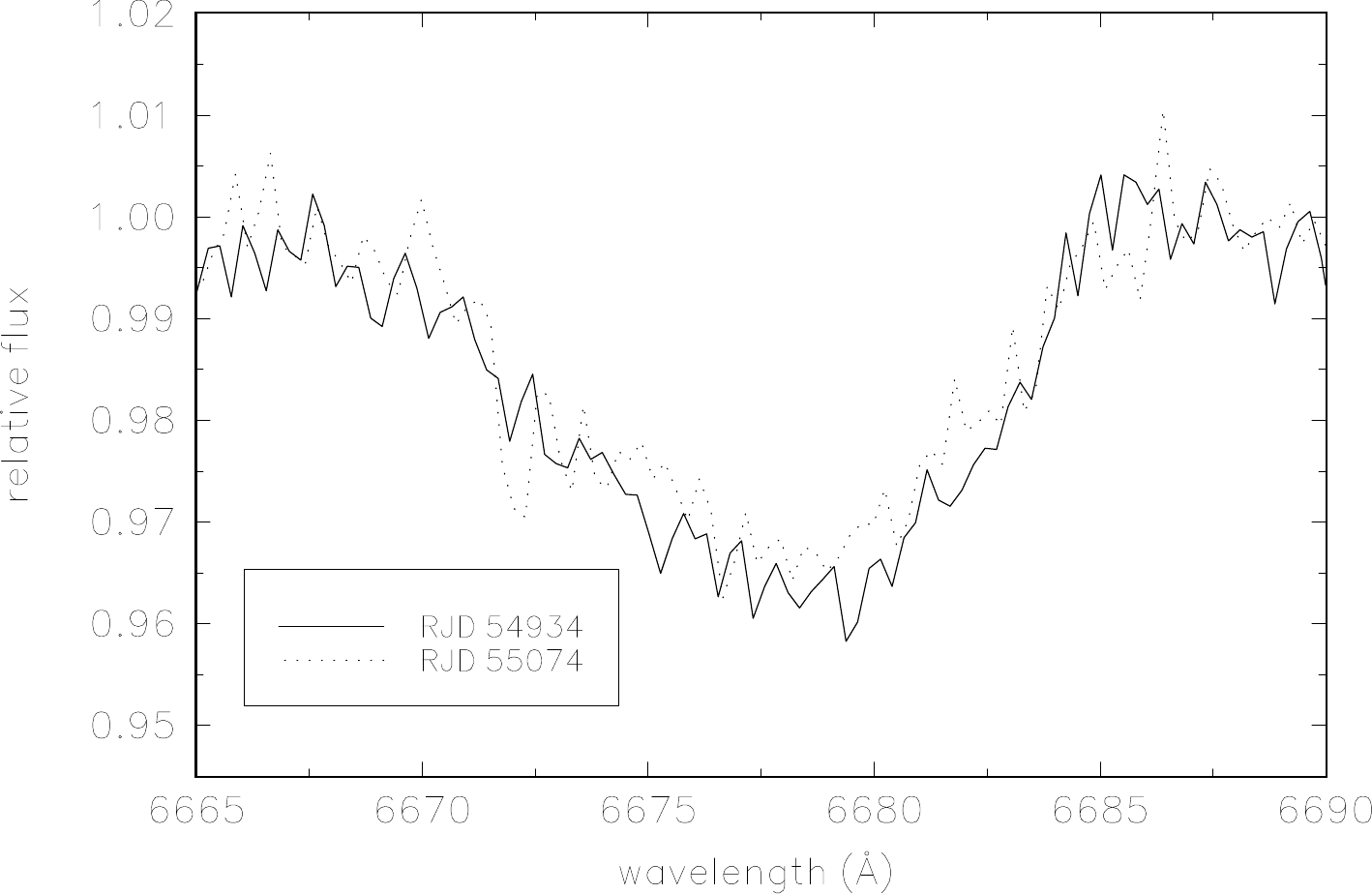}}
\caption{Comparison of the OND spectra from different orbital cycles
in the vicinity of the \he line. Top panel: Spectra from the
elongation with the maximum RV of the primary. Bottom
panel: Two spectra from the other elongation with minimum RV
of the primary. The line asymmetry can clearly be seen to vary
with orbital phase.}\label{minimax}
\end{figure}

\subsection{The nature of the line-profile changes of the \he line}
\citet{vennes}, using phase-averaged spectra, argued that the asymmetry
of the \he line profiles varies with the orbital phase and tentatively
suggested that it might be due to the presence of a weak line of the
secondary component. On the other hand - since the pioneering work
by \citet{wyf79} - it is well known that the hot emission-line stars are
often exhibiting rapid line-profile changes, sometimes in the form of
subfeatures travelling across the line profiles.

To get some idea of which interpretation is the most probable, we inspected
the homogeneous series of the OND spectra. First, we inspected the two  series of spectra taken on RJDs~54296 and 54297, of
0.18 and 0.21~d duration, respectively.
We found no detectable systematic change in these profiles.
On the other hand, the line profiles from different orbital cycles and from
the two opposite elongations clearly demonstrate that the line asymmetry
indeed varies with the orbital phase; see Fig.~\ref{minimax}.
This seems to support the conclusion by \citet{vennes}.

\subsection{\korel disentangling}

\begin{figure}
\centering
\resizebox{\hsize}{!}{\includegraphics{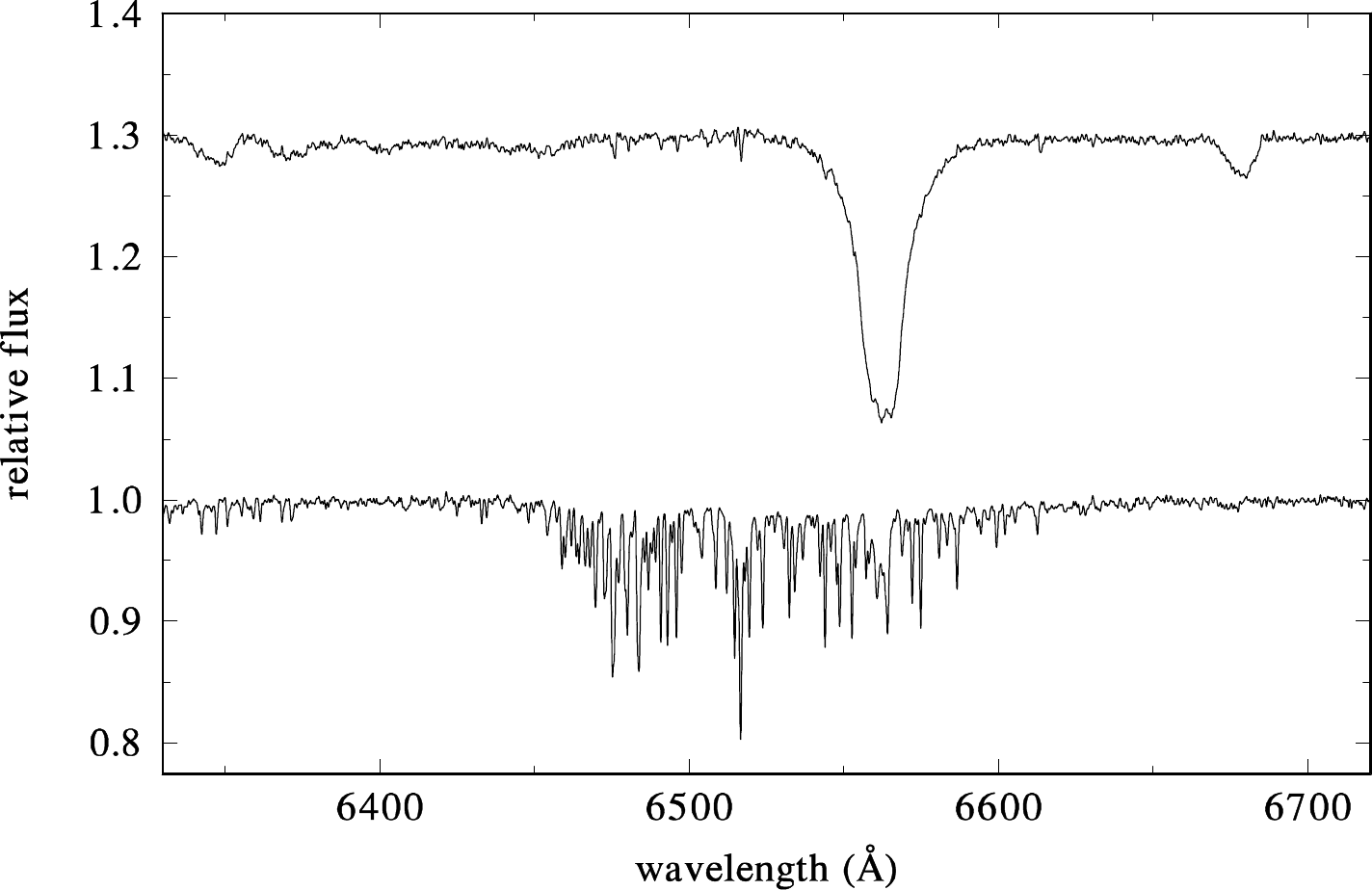}}
\caption{Result of \korel disentangling for 14 OND spectra secured in
2019, when the emission was almost absent. The top panel, shifted
for 0.3 in normalised flux for clarity, shows the disentangled spectrum
of the primary, while the bottom panel shows the disentangled spectrum
of telluric lines.}\label{korel}
\end{figure}

Considering the above finding, it was deemed useful to carry out
an attempt for spectra disentangling. We used the program \korel
\citep{korel} for this purpose. However, with the spectra at hand,
spectra disentangling is a~difficult task. The strongest line at our
disposal, \hae, is not usable because of its secular variations.
There are only 15 OND spectra from the year 2019, when the emission was
almost absent, all of which contain strong water vapour lines and are simply
not numerous enough to detect a weak signal from the secondary.
We are thus left with 107 line profiles of the (relatively weak)
\ion{He}{i}~6678~\AA\ line. We first disentangled the region
6510 - 6539~\AA, which contains strong water vapour lines, to derive
the relative line strengths of the telluric lines for all spectra
\citep{korel2}. These line strengths were then kept fixed
in all consecutive trials.

We then disentangled the whole red spectrum of the primary (6330 - 6720~\AA)
from the epoch without emission. This is shown together with
the disentangled telluric spectrum in Fig.~\ref{korel}. Lines of
\ion{Si}{ii} at 6347 and 6371~\AA,  \ion{Ne}{i}~6402~\AA, \hae, and
\ion{He}{i}~6678~\AA\ are seen there. This disentangled spectrum of
the primary was later used to estimate its radiative properties; see Sect.~\ref{synt}.

\begin{table}
\caption[]{An unconstrained orbital solution for the \he line
derived by \korele.}
\label{korelsol}
\begin{center}
\begin{tabular}{rcccccl}
\hline\hline\noalign{\smallskip}
Element            & Value     \\
\noalign{\smallskip}\hline\noalign{\smallskip}
$P$ (d)               & 69.4212 fixed   \\
$T_{\rm periastr.}$   &55986.733        \\
$e$                   & 0.1077          \\
$\omega$ ($^\circ$)   &265.1            \\
$K_1$ (\ks)           & 8.753           \\
$K_2$ (\ks)           & 86.092          \\
$m_1\sin^3i$ (\Mnom)  & 5.474           \\
$m_2\sin^3i$ (\Mnom)  & 0.557           \\
$\Sigma(O-C)^2$       &  19473.4         \\
No. of RVs            & 102             \\
\noalign{\smallskip}\hline\noalign{\smallskip}
\end{tabular}
\tablefoot{The epoch is in RJD; the sum of squares of residuals is
calculated over the differences between the orbital model and all observed
spectra; see \citet{korel} for details.
\korel does not provide error estimates, so we give all values for
a higher number of digits to prevent round-off errors.}
\end{center}
\end{table}

\begin{figure}
\centering
\resizebox{\hsize}{!}{\includegraphics{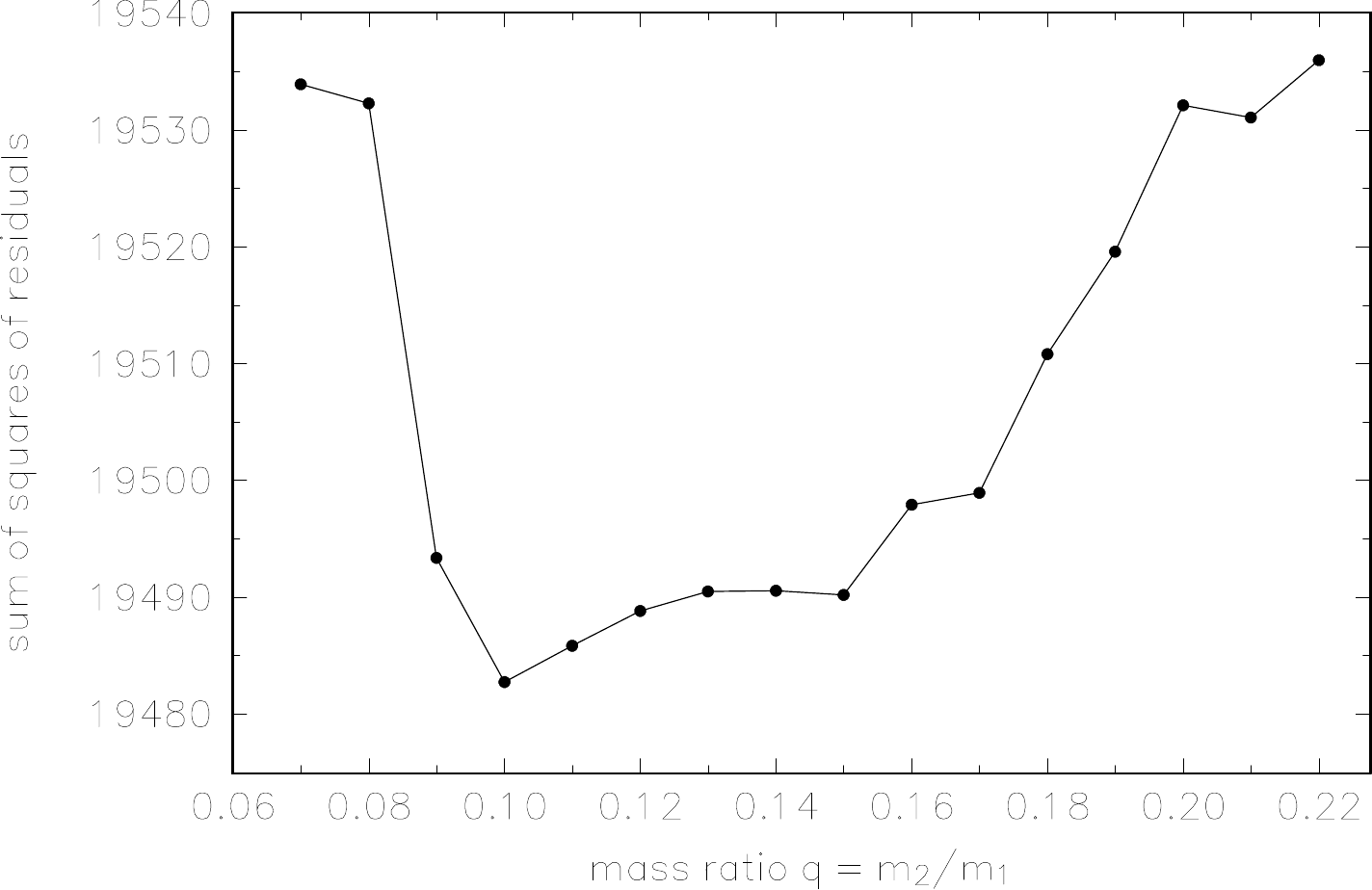}}
\caption{Dependence of the sum of squares of residuals on
the binary mass ratio $q$ derived from the \korel solutions in
the neighbourhood of the \ion{He}{i}~6678~\AA \ line. All elements from our
solution were kept fixed and different fixed values of the mass ratio were
used. The dependence is flat but a clear minimum for mass ratios
between 0.10 and 0.15 is seen.}\label{qmap}
\end{figure}

\begin{figure}
\centering
\resizebox{\hsize}{!}{\includegraphics{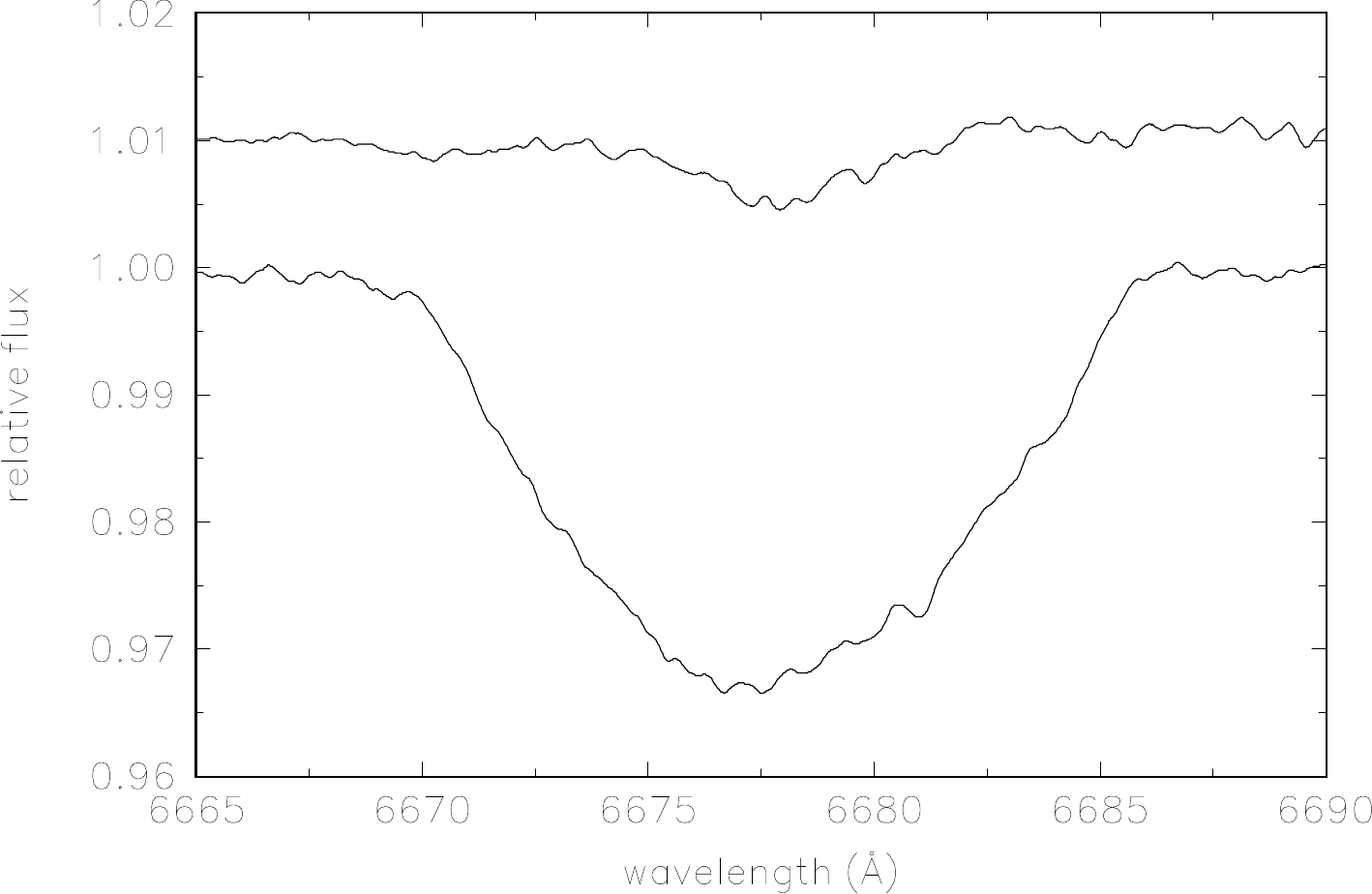}}
\caption{Result of \korel disentangling of both binary components in the
neighbourhood of \ion{He}{i}~6678~\AA \ for all DAO and OND spectra. For
clarity, the disentangled profile of the secondary was shifted up by 0.01.
Both spectra were normalised relative to the joint continuum of
the binary.}\label{he12}
\end{figure}

 For the final attempt, we used only the spectral region from 6665 to
6690~\AA, which contains the \ion{He}{i}~6678~\AA \ line. Noting generally
higher noise levels in the BeSS spectra in this region, we restricted our
analysis to 102 DAO and OND spectra and tried to disentangle the primary
and secondary line profiles, keeping all orbital elements fixed at values
from Table~\ref{orbit}. We derived the goodness-of-fit for a range of
fixed values of the mass ratio as shown in Fig.~\ref{qmap}. It is seen
that there is a clearly defined minimum of the sum of squares of residuals for
the mass ratios between about 0.10 and 0.15. We therefore allowed for
unconstrained solutions, keeping only the orbital period fixed and seeking
the solution with the lowest sum of squares of residuals. This is
presented in Table~\ref{korelsol}.
Disentangled line profiles of the primary and secondary, normalised to the
joint continuum of the binary, are shown in Fig.~\ref{he12}. Measured
RVs of both profiles are close to the systemic velocity of the binary
of about $-16$~\ks, which verifies that the line in the secondary spectrum
is the \ion{He}{i}~6678~\AA \ line and not some other line. We very roughly
estimate its projected rotational velocity $v_2\sin i \sim 100$~\ks.
As pointed out to us by an anonymous referee, \citet{choj2018} reported
a discovery of a hot subdwarf secondary in the binary system HD~55606
with a Be primary and with orbital elements quite similar to \ve: period
93\fd76, semi-amplitudes of 11 and 78 \kms , and a mass ratio of 0.14.
This leads us to a tentative conclusion that the secondary
of \va is  also a hot subdwarf. New high-$S/N$ echelle spectra, covering
long wavelength intervals will be needed to reliably disentangle more
spectral lines of the secondary and for the determination of its
radiative properties. Because of the relatively cooler temperature
(\teff $\sim$15000 -- 16000 K) of the Be star in \va, a hot subdwarf may
be easier to detect in the \va system given the lower contrast ratio
compared to some other Be + hot subdwarf binaries, such as HD~55606
(Be star \teff $\sim$21000 K) and $\varphi$~Per
\citep[Be star \teff $\sim$25500~K;][]{Fremat2005}.

\begin{figure}
\centering
\resizebox{\hsize}{!}{\includegraphics{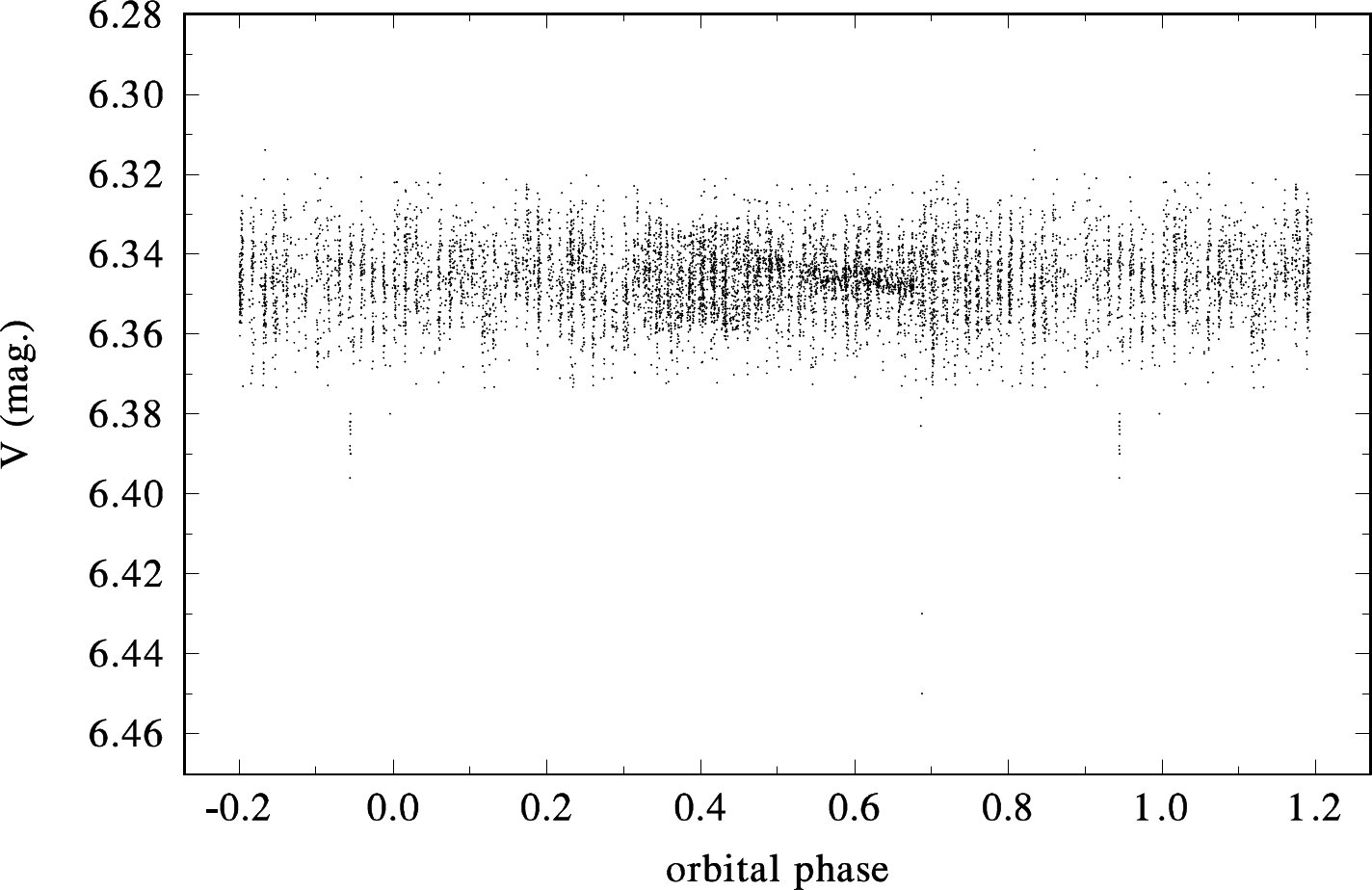}}
\caption{Combined Hvar $V$ photometry, and Hipparcos, KELT,
and TESS photometries transformed to $V$ vs. phase of the 69\fd4212 orbital
period, with phase zero corresponding to a~conjunction with the Be star behind
the secondary according to our orbital
solution. This different phase origin is used for convenience
because a possible primary eclipse should be observed near phase zero
in case of a detached late-type secondary, or near phase 0.5 if
the secondary is a compact subdwarf hotter than the Be primary.}
\label{vfaze}
\end{figure}

\subsection{Orbital light changes?}
 We also investigated possible orbital light variations of \ve. To this
purpose, we used only four sets of well-calibrated observations:
Hvar $V$ magnitude, Hipparcos \hp \ magnitude transformed to $V$,
shifted KELT $R$ magnitude, and shifted TESS magnitude.
We derived robust mean values for the
combined Hvar and Hipparcos transformed $V$ magnitude, and for the KELT
$R$ magnitude and detrended TESS magnitudes, finding  that it is necessary
to subtract 0\m8563 from the KELT data, and add 6\m3465 to the TESS data
to bring them on the scale of the $V$ magnitude. This
procedure seems justified if one notes that the $V-R$ index of \va
is close to zero. The corresponding phase plot is in Fig.~\ref{vfaze}.
We note that there are some transient light decreases, which are  systematic during the corresponding nights of observations, but these
show no clear relation to orbital phases.  The fact that
they were occasionally detected in three considered independent
photometries and also in the DAO photometry confirms their reality.
For the moment, we have no clear physical interpretation for them and
we were unable to confirm any periodicity in their occurrence. They
are obviously quite rare.  We conclude that it is possible to exclude
any detectable binary eclipses in the \va system.

\section{Rapid light changes}\label{rapidvar}

 Two reports of a rapid light variability of \va were published.
\citet{koen2002} reported a sinusoidal variation with a period of
0\fd5592278 and an amplitude of 0\m0088 from their analysis of \hp
\ photometry. \citet{bartz2017} reported a period of 4\fd66436 from
the analysis of numerous KELT photometric observations.

\begin{figure}
\centering
\resizebox{\hsize}{!}{\includegraphics{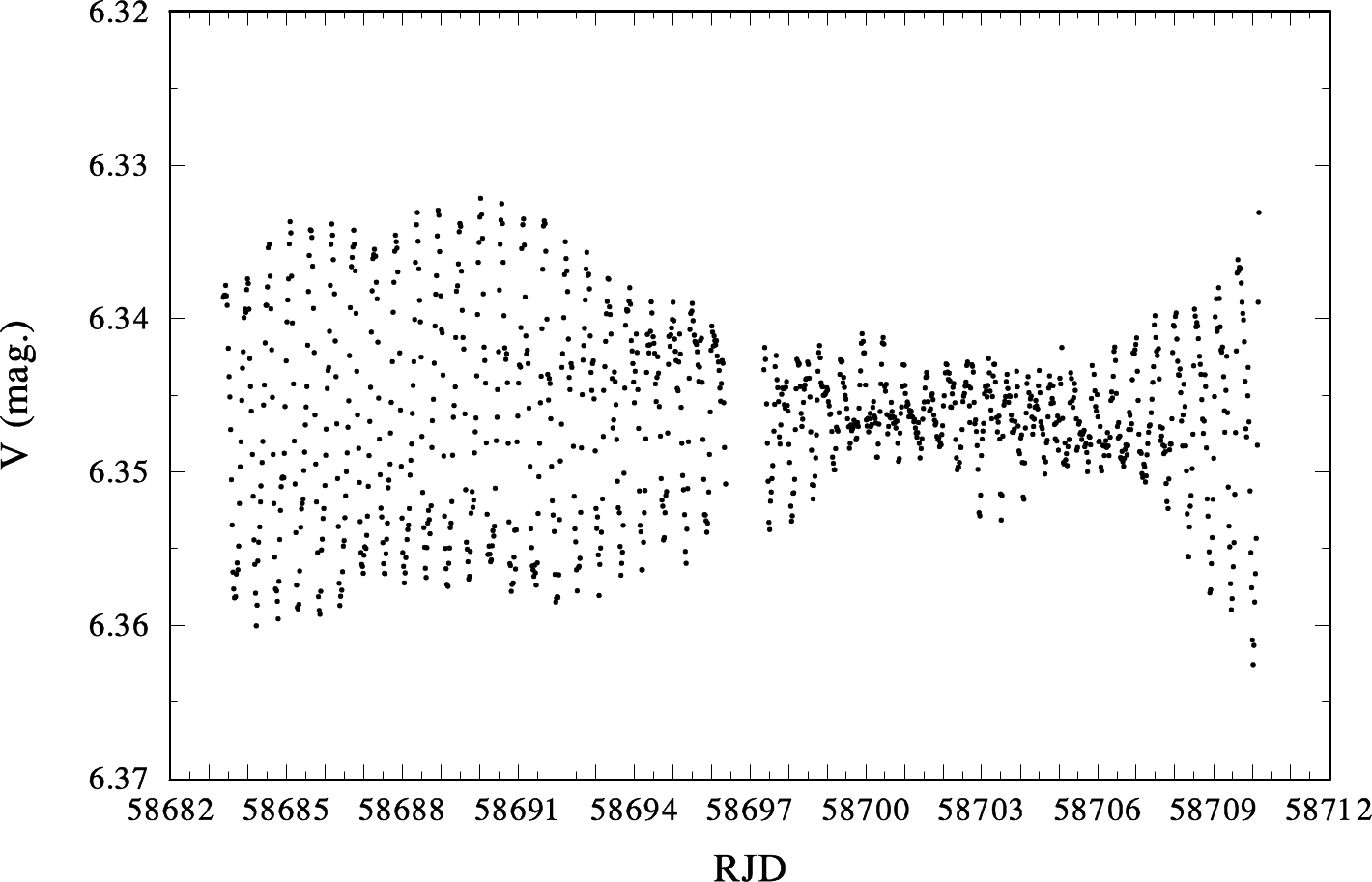}}
\caption{Time plot of accurate TESS photometry of \ve.
It is seen that the amplitude of the 0\fd559 period varies with time.}
\label{tesstime}
\end{figure}

\begin{figure}
\centering
\resizebox{\hsize}{!}{\includegraphics{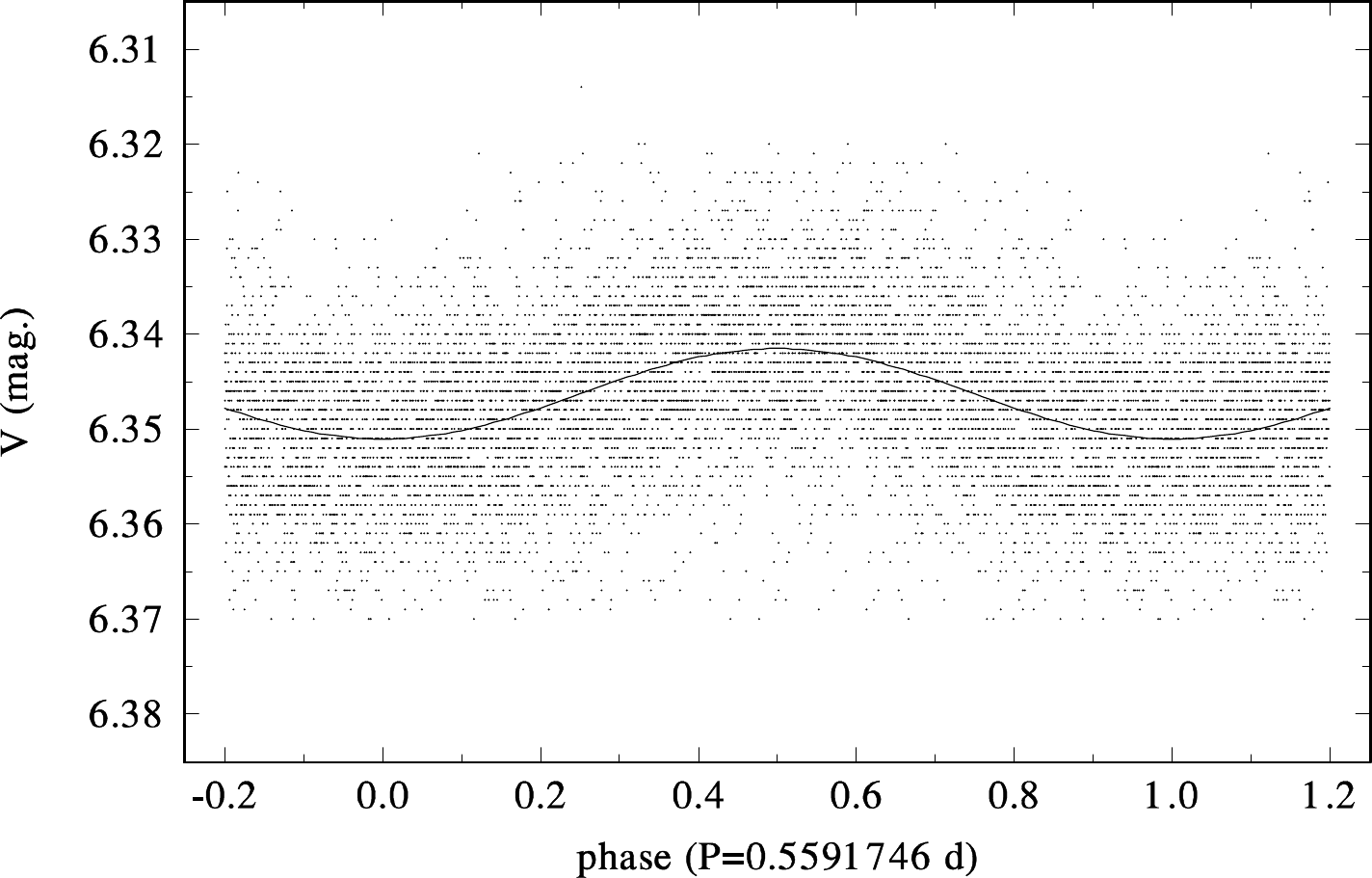}}
\caption{Periodic rapid light variations based on calibrated
Hvar $V$ photometry, \hp \ photometry transformed to $V$ magnitude,
KELT $R$ photometry decreased by 0\m8563, and detrended TESS broadband
magnitudes increased by 6\m3465 to shift them to the range of $V$
photometry. All observations from rapid light decreases that were fainter
than 6\m37 were omitted. All data are plotted with the ephemeris
derived from a sinusoidal fit
$T_{\rm min.V}={\rm RJD}~53000.2951(48)+0\fd55916706(83)\cdot E$.
The fitted sinusoid with a semi-amplitude of 0\m00476(13) is
shown as a solid line.}\label{faze559}
\end{figure}

Omitting the DAO data, which were obtained relative to another Be variable,
V395 Vul = 12 Vul, we analysed as before the combined set of the transformed
Hipparcos \hp \ photometry, the Hvar $V$ magnitude observations, KELT
$R$ photometry decreased arithmetically by 0\m8563, and TESS detrended photometry increased by 6\m3465 to shift them to
the range of the $V$ magnitude. For the search for rapid variability,
we omitted all observations fainter than 6\m37 to exclude
the non-periodic occasional light decreases mentioned above. A period search in this combined data set returned a period of 0\fd55917. We carried out
a~sinusoidal fit to obtain the most accurate value for this period,
0\fd55916706$\pm$0\m00000083. The full amplitude of this variability is
0\m00476$\pm$0\m00013. However, since different filters are used in
the various photometric datasets, and especially since the amplitude is
clearly variable (as seen in the TESS data in Fig.~\ref{tesstime}), this
amplitude is necessarily approximate. The corresponding sinusoidal 0\fd559
light curve is shown in Fig.~\ref{faze559}. Its scatter is higher than
observational errors and we attribute this to the above mentioned variations
in the light-curve amplitude and to probable secular changes in the
accurate value of the 0\fd559 period itself. The signal of the rapid light variability is clearly more complex than a simple sinusoid at a single frequency, as evidenced by the apparent amplitude variation in the TESS light curve, and the presence of many harmonics in the Fourier transform of the TESS data. A more detailed investigation
of the nature of rapid variations of \va is beyond the scope of this
paper and will be published elsewhere. We note that the mean
0\fd559 period could be interpreted as the rotational period of the primary,
which we estimate in the following section to be about 0\fd5.

We would also like to mention that no convincing
light curve could be found in the combined photometry for periods
near  4\fd664,  which was advocated by \citet{bartz2017}.
A re-analysis of the KELT photometry convincingly demonstrates that
the strongest signal is at 0\fd559, and that the 4\fd664 period is
an~alias of the fast period with the daily sampling, i.e.,
frequency (4.664)$^{-1} \sim$ 2 cycles d$^{-1}$ - (0.5592)$^{-1}$.

\section{Probable binary properties}
Since we still consider the detection of the \he line of the secondary
as tentative, we investigate the probable range of the basic physical
properties along two lines using the observed and deduced properties of the
primary only, and using the estimates of binary masses from the \korel
solution.

\subsection{Radiative properties of the primary}\label{synt}
\citet{vennes} estimated the radiative properties of the primary of \va
in several ways. Using the spectral energy distribution (SED) from the UV to IR wavelengths, these latter authors obtained \tef = $14400\pm800$~K and
$E(\bv)=0$\m069\p0\m030. We note that for that particular study, the authors collected
photometry from various sources and from different epochs. They also
considered both determinations of the Hipparcos parallax $p=0\farcs00429\pm0\farcs00076$ \citep{esa97}, and
$p=0\farcs00281\pm0\farcs00048$ \citep{leeuw2007a,leeuw2007b}.
Fitting the high-dispersion Balmer-line profiles with the B-star model grid \citep{bgrid}, they arrived at \tef = $15600\pm200$~K, and \lgg = $3.75\pm0.02$ [cgs], and proposed the spectral class B4-5III-IVe.

To determine the radiative properties of the primary, we used the Python program \pyt \citep{jn2016}, which interpolates in a pre-calculated grid of synthetic spectra. Using a set of observed spectra, \pyt tries to find the optimal fit between the observed and interpolated model spectra and returns the
best-fit values of \teff, \vsin and \lgg.

\begin{figure}
\centering
\resizebox{\hsize}{!}{\includegraphics{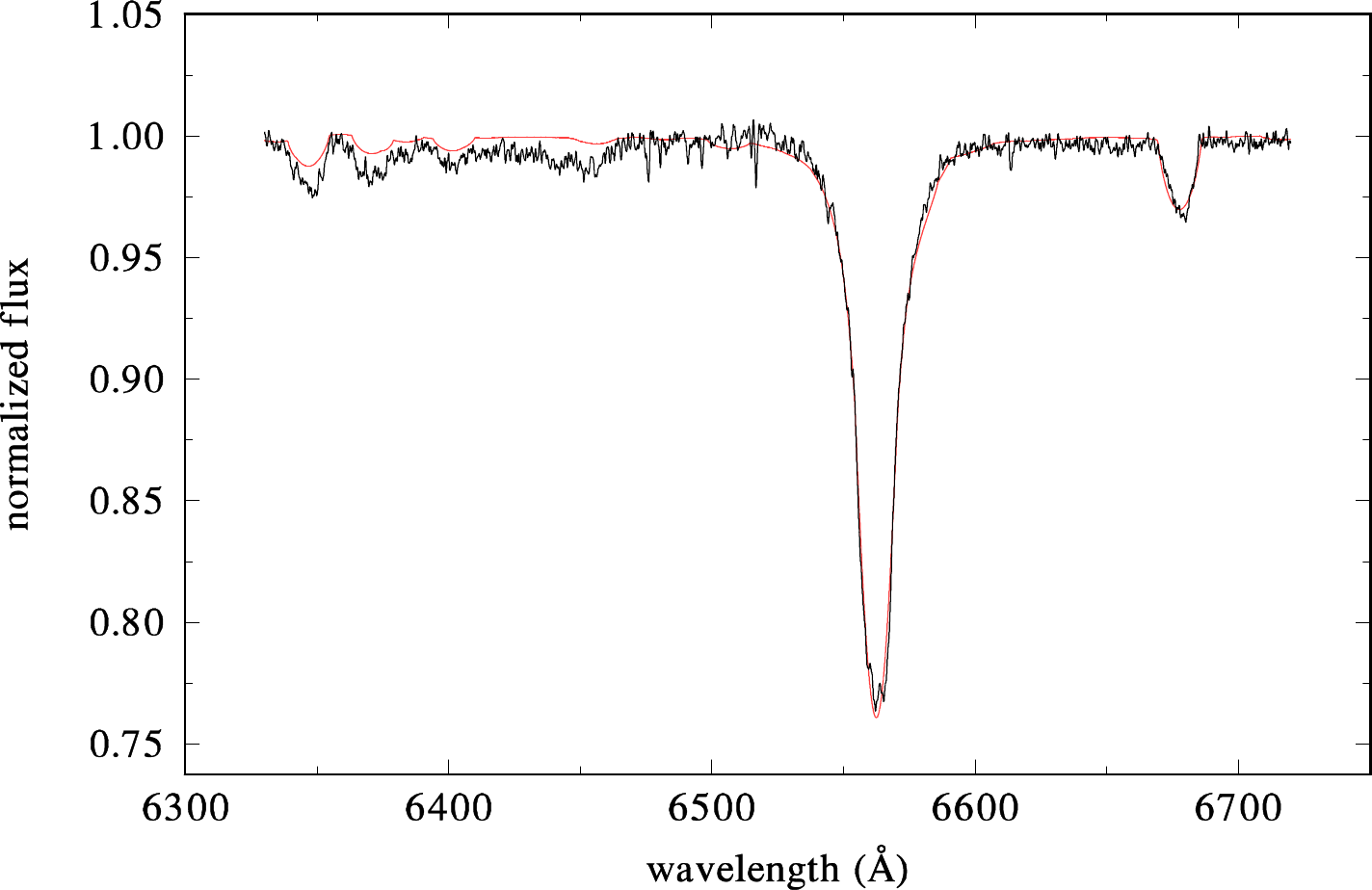}}
\resizebox{\hsize}{!}{\includegraphics{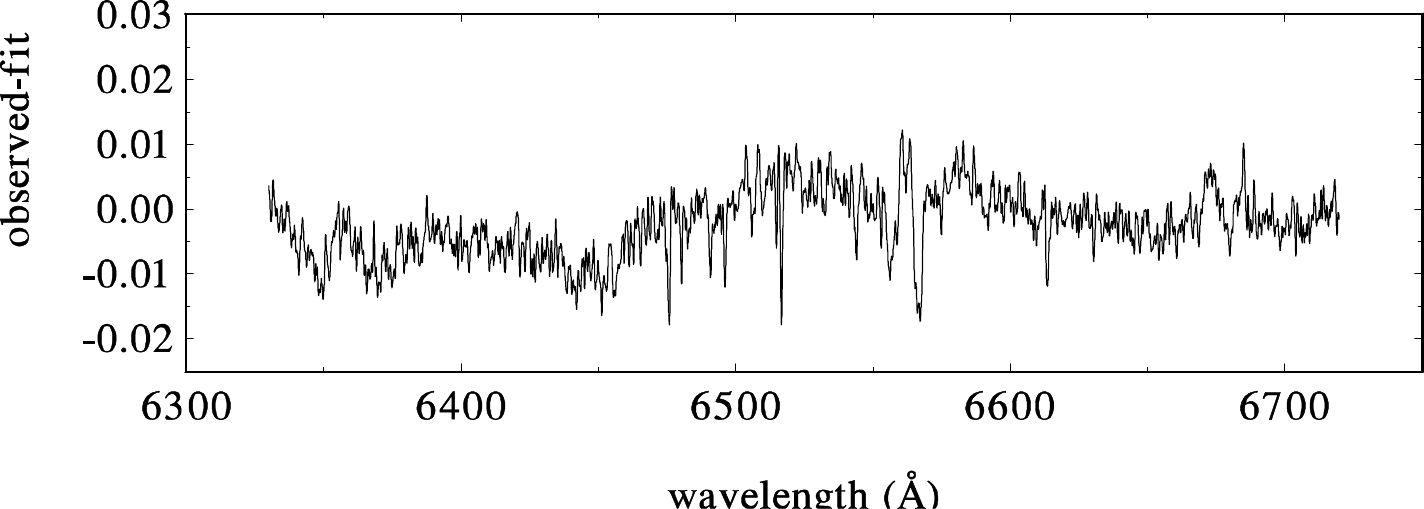}}
\caption{Comparison of the disentangled red spectrum of the primary with an interpolated synthetic spectrum for \tef = 14910~K,
\lgg = 3.55, and \vsin = 376~\ks.
The residuals in the sense observed minus synthetic spectra are also shown in
the bottom panel.}
\label{syn}
\end{figure}
\begin{table}
\caption{Two distinct determinations of the radiative properties of the primary with \pyt and MCMC analysis.}
\label{synpar}
\begin{center}
\begin{tabular}{rcccccl}
\hline\hline\noalign{\smallskip}
Element         & Pollux ATLAS12 models  & B-star grid \\
\noalign{\smallskip}\hline\noalign{\smallskip}
\tef (K)        & 14910 & 15750\\
\lgg [cgs]      & 3.55  & 3.61\\
\vsin (\ks)     & 376   & 369 \\
\noalign{\smallskip}\hline\noalign{\smallskip}
\end{tabular}
\end{center}
\end{table}

In our particular application, two different grids of spectra were used:
a B-star grid of line-blanketed NLTE models \citep{bgrid},
and the Pollux database \citep{pala2010}: the program fitted
the disentangled spectrum of the primary from 2019 spectra
(without the trace of emission) over the whole wavelength range from 6330 to
6720~\AA. We then ran a Markov chain Monte Carlo (MCMC) simulation
using {\tt emcee} Python library by~\citet{fore2013}. \footnote{The library is
available through GitHub~\url{https://github.com/dfm/emcee.git} and its
thorough description is at~\url{http://dan.iel.fm/emcee/current/}.}
We ran 450 iterations and to our surprise, this revealed two distinct, parallel
 solutions. These are given in Table~\ref{synpar} and the fit
for the cooler \teff, found originally from \pyt solution,
is shown in Fig.~\ref{syn}.  It was pointed to us by our colleague
M.~Bro\v{z} that the two separate solutions are indicative of mutual
differences between the B-star and Pollux grid. The lowest \tef of
the B-star grid is 15000~K. Therefore, the two distinct solutions correspond to
the two sets of synthetic spectra. Considering
that the rapid stellar rotation of the primary can also affect the observed
spectrum, we do not think that any more sophisticated techniques
would provide us with real error estimates of the radiative properties
of the primary. These are hidden in systematic differences in the different
model spectra and their uncertainties. We therefore think that the difference
between the two solutions represents a good guess of what the real
uncertainties could be. We note that an estimate of \tef by
\citet{vennes} based on the SED also falls
within this range.

\subsection{Luminosity, mass, and radius of the primary}
\begin{figure}
\centering
\resizebox{\hsize}{!}{\includegraphics{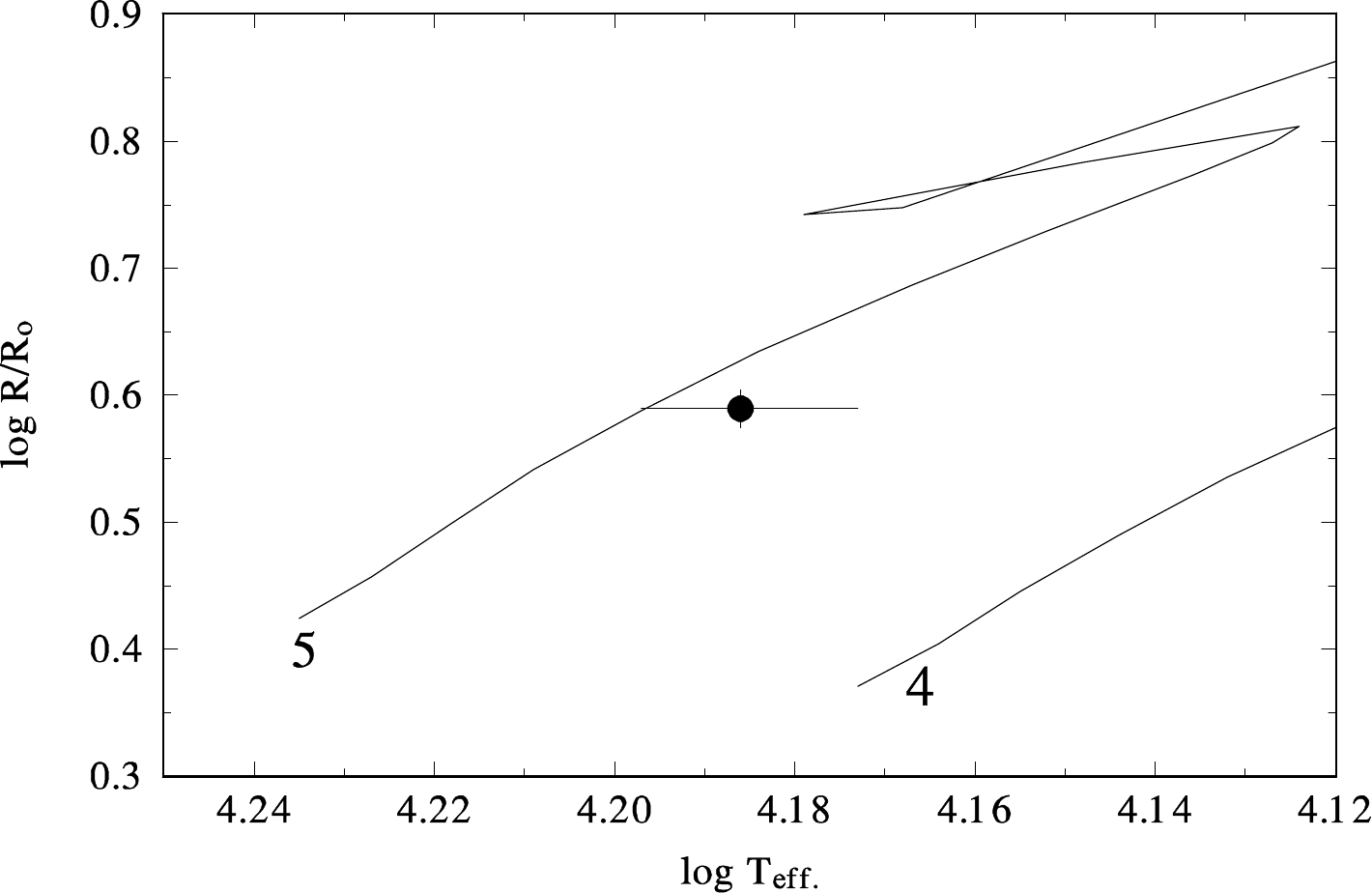}}
\caption{Comparison of the range of possible values of the radius of
 the primary, which we estimated for the two distinct values of \teff,
the range of dereddened $V$ magnitude, and the Gaia parallax.
We note that the two evolutionary tracks are labelled by stellar mass.
}
\label{logr}
\end{figure}

A standard dereddening of the mean calibrated \ubv\ magnitudes
from more recent Hvar observations,

\smallskip
\centerline{$V=6$\m337, \bv=$-0$\m093, and \ub=$-0$\m539,}

\noindent gives

\smallskip
$V_0=6$\m133-6\m153, (\bv)$_0=-0$\m157, (\ub)$_0=-0$\m585,
\smallskip
\noindent with E(\bv)=0\m064. The range in the magnitude corresponds to
the adopted range of $A_{\rm V}$ from 2\m9 to 3\m2.

Using the accurate Gaia Data Release 2 parallax \citep{gaia1, gaia3},

\centerline{$p=0\farcs003473\pm0\farcs000048,$}

\noindent the range of $V_0$ and both possible values of \teff,
we obtain the primary radius $R$ = 3.75 - 4.02~\Rnom, which is notably
smaller than that estimated by \citet{vennes}. The corresponding range
of the absolute magnitude $M_{\rm bol}$ is from $-2$\m35 to $-2$\m52.

In Fig.~\ref{logr} we compare our radii with the evolutionary tracks
of \citet{schaller92} in the $\log~R$ versus $\log~T_{\rm eff.}$ diagram.
From this comparison, we calculate the probable mass of the primary to be $m_1=4.8\pm0.1$~\Mnom, which is quite plausible for a~B4-5IV star.

In passing we note that our values of the mass and radius would imply
\lgg~$\sim3.9$ [cgs]. This is higher than the values of $\sim3.6$ or
$\sim3.75$ [cgs] estimated by \citet{vennes}. Since the Balmer-line profiles
are the most sensitive to \lgge, we propose that they have been affected
by some amount of emission from circumstellar matter throughout
the whole documented spectral history of the star.

\begin{table}
\caption{Probable mass ratio $q=m_2/m_1$ and the mass of the secondary
estimated for several orbital inclinations $i$, $m_1=4.8$~\Mnom\ , and our orbital solution.}
\label{q}
\begin{center}
\begin{tabular}{cccccl}
\hline\hline\noalign{\smallskip}
 $i$    & $m_2$     & $q$ \\
(degs.)&(\Mnom)    \\
\noalign{\smallskip}\hline\noalign{\smallskip}
 90.0 &   0.5180 &   0.1079\\
 80.0 &   0.5266 &   0.1097\\
 70.0 &   0.5538 &   0.1154\\
 60.0 &   0.6047 &   0.1260\\
 50.0 &   0.6908 &   0.1439\\
 40.0 &   0.8379 &   0.1746\\
 30.0 &   1.1118 &   0.2316\\
\noalign{\smallskip}\hline\noalign{\smallskip}
\end{tabular}
\end{center}
\end{table}

\subsection{Probable physical properties of the binary system}

\begin{table}
\caption{Masses of the binary components estimated for several possible
orbital inclinations from the \korel solution of Table~\ref{korelsol}.}
\label{masses}
\begin{center}
\begin{tabular}{cccccl}
\hline\hline\noalign{\smallskip}
 $i$   &   $m_1$ &  $m_2$  \\
(degs.)&(\Mnom)  &(\Mnom)  \\
\noalign{\smallskip}\hline\noalign{\smallskip}
 90.0 &   5.47   &   0.56  \\
 80.0 &   5.73   &   0.58  \\
 70.0 &   6.60   &   0.67  \\
 58.0 &   8.98   &   0.91  \\
\noalign{\smallskip}\hline\noalign{\smallskip}
\end{tabular}
\end{center}
\end{table}

If we adopt the mass of 4.8~\Mnom\ and our orbital solution, we
can estimate the mass ratio and the mass of the secondary for a few
different orbital inclinations. These are summarised in Table~\ref{q}.
If we assume that the axis of orbital revolution and the axis of the rotation of the primary are parallel, we can set a lower limit for the orbital
inclination. Adopting the Roche model approximation, we have
$R_{\rm equator}=1.5R_{\rm pole}$. In the limiting case of break-up rotation
of the primary, we can roughly approximate the area, from which we
receive the stellar flux of the primary (estimated from dereddened observed $V$ and parallax) by an effective radius $R_{ef}$, for which we have
(assuming the area of an ellipse),
\begin{equation}
\pi R_{\rm pole}R_{\rm equator}=\pi R^2_{ef} \rightarrow
R_{\rm equator}=\sqrt{3/2}R_{ef}
.\end{equation}
\noindent For $R_{ef}\approx3.9$~\Rnom, $R_{\rm equator}\approx4.8$~\Rnom.
Consequently, for critical rotation at the equator, we obtain $\approx440$~\ks,
which implies that the inclination must be higher then about $58^\circ$.
It is notable that this last estimate agrees well with that of
\citet{vennes}, although we used different values of basic
physical properties of the system. Finally, we note that the probable
rotational period of the primary should be between about 0\fd45 and
0\fd53 for the range of inclinations from $58^\circ$ to $90^\circ$.

  Using a different approach, \citet{kervella2019} estimated the mass and
radius of the \va primary as

\centerline{$m=5.00\pm0.25$~\ms\  and $R=4.64\pm0.23$~\rs.}

\noindent If we assume that the 0\fd5592 period is the rotational
period of the primary and adopt our \vsin = 370~\ks,
we find $R_{\rm equator}=4.09 - 4.82$~\Rnom \ for the same range of inclinations
as above.

 Finally, if we use the \korel solution shown in Table~\ref{korelsol},
we obtain possible masses of both components as a function of orbital
inclination. These are listed for several plausible values of the
inclination in Table~\ref{masses}. From this, we conclude that
the true orbital inclination should be higher than some $70^\circ$.

\section{Conclusions}
Our analysis confirms the binary nature of \va and a low mass for
its companion. We underline that our estimates are preliminary due
to all the problems we have discussed above. The secondary cannot be
a cool Roche-lobe-filling object because, were this the case, its spectral lines would
have been detected in the observed spectra \citep[cf, e.g.][]{zarf30}.
In principle, it could be a normal star of a late spectral class but
the existence of such companions to Be stars has not yet been
proven. It is much more probable that this secondary is a hot subdwarf like that
observed by \citet{gies98} for $\varphi$~Per;  see also the recent
search for such systems in the existing IUE spectra \citep{wang2018}
and the detection of a hot subdwarf for HD~55606 \citep{choj2018}.
To the best of our knowledge, there are no high-dispersion far-UV
spectra of \ve. In order to make progress in the
understanding of this system, we feel it necessary to
first of all obtain a series of high-S/N far-UV spectra over the whole
69.4~d period and search for the possible presence of lines of a hot
subdwarf secondary. We believe it would furthermore be helpful to monitor the \ha profile of the star from time to time and
to obtain another series of high-resolution high-S/N optical echelle
spectra, when the star appears again as a~normal rapidly rotating
B star, and to carry out a much more accurate orbital analysis and
line-profile modelling.
 Lastly, several whole-night series of high-S/N spectral observations
could be used to check for the presence of mild rapid line-profile variations,
possibly related to the rapid periodic light changes.

\medskip
\begin{acknowledgements}
We acknowledge the use of the program \pyt by J.~Nemravov\'a, and
the programs \fotel and \korel written by P.~Hadrava. An illuminating discussion
on the problems of model spectra with M.~Bro\v{z} is also appreciated. We thank
P.~Hadrava, A.~Kawka, D.~Kor\v{c}\'akov\'a, M.~Kraus, J.~Kub\'at,
B.~Ku\v{c}erov\'a, P.~Nem\'eth, M.~Netolick\'y, J.~Polster, S.~Vennes,
and V.~Votruba, who obtained a number of \ond\ spectra used in this study.
A.~Opli\v{s}tilov\'a and K.~Vitovsk\'y helped to secure some
\ubvr \ observations at Hvar.
This work has made use of the BeSS database, operated at LESIA,
Observatoire de Meudon, France: http://basebe.obspm.fr and we thank
the folowing amateur observers, who contributed their spectra:
V.~Desneoux, A.~Favaro, O.~Garde, K.~Graham, F.~Houpert, and O.~Thizy.
Useful suggestions and critical remarks of an anonymous referee
helped to improve the presentation and {some analyses} and
are gratefully acknowledged.
P.H. was supported by the Czech Science Foundation grant GA19-01995S.
H.B. acknowledges financial support from Croatian Science Foundation under
the project 6212 ``Solar and Stellar Variability".
J.L. and A.H. were supported by student research grants of the faculty
of Mathematics and Physics of Charles university.
J.L.-B. acknowledges support from FAPESP (grant 2017/23731-1).
This project makes use of data from the KELT survey, including support from
The Ohio State University, Vanderbilt University, and Lehigh University,
along with the KELT follow-up collaboration.
The following internet-based resources were
consulted: the SIMBAD database and the VizieR service operated at
CDS, Strasbourg, France; and the NASA's Astrophysics Data System
Bibliographic Services. This work has made use of data from
the European Space Agency (ESA) mission Gaia
(\url{https://www.cosmos.esa.int/gaia}), processed by the Gaia
Data Processing and Analysis Consortium (DPAC;
\url{https://www.cosmos.esa.int/web/gaia/dpac/consortium}).
Funding for the DPAC has been provided by national institutions,
in particular the institutions participating in
the Gaia Multilateral Agreement.
This research made use of Lightkurve, a Python package for Kepler and TESS data analysis \citep{Lightkurve}.
\end{acknowledgements}

\bibliographystyle{aa}
\bibliography{7vul}

\begin{thebibliography}{59}
\expandafter\ifx\csname natexlab\endcsname\relax\def\natexlab#1{#1}\fi

\bibitem[{{Baumgardt}(1998)}]{baum98}
{Baumgardt}, H. 1998, \aap, 340, 402

\bibitem[{{Chambliss}(1977)}]{chamb77}
{Chambliss}, C.~R. 1977, Information Bulletin on Variable Stars, 1233, 1

\bibitem[{{Chojnowski} {et~al.}(2018){Chojnowski}, {Labadie-Bartz}, {Rivinius},
  {Gies}, {Panoglou}, {Borges Fernandes}, {Wisniewski}, {Whelan}, {Mennickent},
  {McMillan}, {Dembicky}, {Gray}, {Rudyk}, {Stringfellow}, {Lester},
  {Hasselquist}, {Zharikov}, {Levenhagen}, {Souza}, {Leister}, {Stassun},
  {Siverd}, \& {Majewski}}]{choj2018}
{Chojnowski}, S.~D., {Labadie-Bartz}, J., {Rivinius}, T., {et~al.} 2018, \apj,
  865, 76

\bibitem[{{Crawford} {et~al.}(1971){Crawford}, {Barnes}, \& {Golson}}]{craw71}
{Crawford}, D.~L., {Barnes}, J.~V., \& {Golson}, J.~C. 1971, \aj, 76, 1058

\bibitem[{{Delplace}(1970)}]{delplace70}
{Delplace}, A.~M. 1970, \aap, 7, 68

\bibitem[{{Doazan} {et~al.}(1982{\natexlab{a}}){Doazan}, {Harmanec},
  {Koubsk\'y}, {Krpata}, \& {\v{Z}d\'arsk\'y}}]{doazan82a}
{Doazan}, V., {Harmanec}, P., {Koubsk\'y}, P., {Krpata}, J., \&
  {\v{Z}d\'arsk\'y}, F. 1982{\natexlab{a}}, \aaps, 50, 481

\bibitem[{{Doazan} {et~al.}(1982{\natexlab{b}}){Doazan}, {Harmanec},
  {Koubsk\'y}, {Krpata}, \& {\v{Z}d\'arsk\'y}}]{doazan82b}
{Doazan}, V., {Harmanec}, P., {Koubsk\'y}, P., {Krpata}, J., \&
  {\v{Z}d\'arsk\'y}, F. 1982{\natexlab{b}}, \aap, 115, 138

\bibitem[{{Foreman-Mackey} {et~al.}(2013){Foreman-Mackey}, {Hogg}, {Lang}, \&
  {Goodman}}]{fore2013}
{Foreman-Mackey}, D., {Hogg}, D.~W., {Lang}, D., \& {Goodman}, J. 2013, \pasp,
  125, 306

\bibitem[{{Fr{\'e}mat} {et~al.}(2005){Fr{\'e}mat}, {Zorec}, {Hubert}, \&
  {Floquet}}]{Fremat2005}
{Fr{\'e}mat}, Y., {Zorec}, J., {Hubert}, A.~M., \& {Floquet}, M. 2005, \aap,
  440, 305

\bibitem[{{Gaia Collaboration} {et~al.}(2018){Gaia Collaboration}, {Brown},
  {Vallenari}, {Prusti}, {de Bruijne}, {Babusiaux}, {Bailer-Jones}, {Biermann},
  {Evans}, {Eyer}, {Jansen}, {Jordi}, {Klioner}, {Lammers}, {Lindegren},
  {Luri}, {Mignard}, {Panem}, {Pourbaix}, {Randich}, {Sartoretti}, {Siddiqui},
  {Soubiran}, {van Leeuwen}, {Walton}, {Arenou}, {Bastian}, {Cropper},
  {Drimmel}, {Katz}, {Lattanzi}, {Bakker}, {Cacciari}, {Casta{\~n}eda},
  {Chaoul}, {Cheek}, {De Angeli}, {Fabricius}, {Guerra}, {Holl}, {Masana},
  {Messineo}, {Mowlavi}, {Nienartowicz}, {Panuzzo}, {Portell}, {Riello},
  {Seabroke}, {Tanga}, {Th{\'e}venin}, {Gracia-Abril}, {Comoretto},
  {Garcia-Reinaldos}, {Teyssier}, {Altmann}, {Andrae}, {Audard},
  {Bellas-Velidis}, {Benson}, {Berthier}, {Blomme}, {Burgess}, {Busso},
  {Carry}, {Cellino}, {Clementini}, {Clotet}, {Creevey}, {Davidson}, {De
  Ridder}, {Delchambre}, {Dell'Oro}, {Ducourant},
  {Fern{\'a}ndez-Hern{\'a}ndez}, {Fouesneau}, {Fr{\'e}mat}, {Galluccio},
  {Garc{\'\i}a-Torres}, {Gonz{\'a}lez-N{\'u}{\~n}ez}, {Gonz{\'a}lez-Vidal},
  {Gosset}, {Guy}, {Halbwachs}, {Hambly}, {Harrison}, {Hern{\'a}ndez},
  {Hestroffer}, {Hodgkin}, {Hutton}, {Jasniewicz}, {Jean-Antoine-Piccolo},
  {Jordan}, {Korn}, {Krone-Martins}, {Lanzafame}, {Lebzelter}, {L{\"o}ffler},
  {Manteiga}, {Marrese}, {Mart{\'\i}n-Fleitas}, {Moitinho}, {Mora}, {Muinonen},
  {Osinde}, {Pancino}, {Pauwels}, {Petit}, {Recio-Blanco}, {Richards},
  {Rimoldini}, {Robin}, {Sarro}, {Siopis}, {Smith}, {Sozzetti}, {S{\"u}veges},
  {Torra}, {van Reeven}, {Abbas}, {Abreu Aramburu}, {Accart}, {Aerts},
  {Altavilla}, {{\'A}lvarez}, {Alvarez}, {Alves}, {Anderson}, {Andrei},
  {Anglada Varela}, {Antiche}, {Antoja}, {Arcay}, {Astraatmadja}, {Bach},
  {Baker}, {Balaguer-N{\'u}{\~n}ez}, {Balm}, {Barache}, {Barata}, {Barbato},
  {Barblan}, {Barklem}, {Barrado}, {Barros}, {Barstow}, {Bartholom{\'e}
  Mu{\~n}oz}, {Bassilana}, {Becciani}, {Bellazzini}, {Berihuete}, {Bertone},
  {Bianchi}, {Bienaym{\'e}}, {Blanco-Cuaresma}, {Boch}, {Boeche}, {Bombrun},
  {Borrachero}, {Bossini}, {Bouquillon}, {Bourda}, {Bragaglia}, {Bramante},
  {Breddels}, {Bressan}, {Brouillet}, {Br{\"u}semeister}, {Brugaletta},
  {Bucciarelli}, {Burlacu}, {Busonero}, {Butkevich}, {Buzzi}, {Caffau},
  {Cancelliere}, {Cannizzaro}, {Cantat-Gaudin}, {Carballo}, {Carlucci},
  {Carrasco}, {Casamiquela}, {Castellani}, {Castro-Ginard}, {Charlot},
  {Chemin}, {Chiavassa}, {Cocozza}, {Costigan}, {Cowell}, {Crifo}, {Crosta},
  {Crowley}, {Cuypers}, {Dafonte}, {Damerdji}, {Dapergolas}, {David}, {David},
  {de Laverny}, {De Luise}, {De March}, {de Martino}, {de Souza}, {de Torres},
  {Debosscher}, {del Pozo}, {Delbo}, {Delgado}, {Delgado}, {Di Matteo},
  {Diakite}, {Diener}, {Distefano}, {Dolding}, {Drazinos}, {Dur{\'a}n},
  {Edvardsson}, {Enke}, {Eriksson}, {Esquej}, {Eynard Bontemps}, {Fabre},
  {Fabrizio}, {Faigler}, {Falc{\~a}o}, {Farr{\`a}s Casas}, {Federici},
  {Fedorets}, {Fernique}, {Figueras}, {Filippi}, {Findeisen}, {Fonti},
  {Fraile}, {Fraser}, {Fr{\'e}zouls}, {Gai}, {Galleti}, {Garabato},
  {Garc{\'\i}a-Sedano}, {Garofalo}, {Garralda}, {Gavel}, {Gavras}, {Gerssen},
  {Geyer}, {Giacobbe}, {Gilmore}, {Girona}, {Giuffrida}, {Glass}, {Gomes},
  {Granvik}, {Gueguen}, {Guerrier}, {Guiraud}, {Guti{\'e}rrez-S{\'a}nchez},
  {Haigron}, {Hatzidimitriou}, {Hauser}, {Haywood}, {Heiter}, {Helmi}, {Heu},
  {Hilger}, {Hobbs}, {Hofmann}, {Holland}, {Huckle}, {Hypki}, {Icardi},
  {Jan{\ss}en}, {Jevardat de Fombelle}, {Jonker}, {Juh{\'a}sz}, {Julbe},
  {Karampelas}, {Kewley}, {Klar}, {Kochoska}, {Kohley}, {Kolenberg},
  {Kontizas}, {Kontizas}, {Koposov}, {Kordopatis}, {Kostrzewa-Rutkowska},
  {Koubsky}, {Lambert}, {Lanza}, {Lasne}, {Lavigne}, {Le Fustec}, {Le
  Poncin-Lafitte}, {Lebreton}, {Leccia}, {Leclerc}, {Lecoeur-Taibi},
  {Lenhardt}, {Leroux}, {Liao}, {Licata}, {Lindstr{\o}m}, {Lister}, {Livanou},
  {Lobel}, {L{\'o}pez}, {Managau}, {Mann}, {Mantelet}, {Marchal}, {Marchant},
  {Marconi}, {Marinoni}, {Marschalk{\'o}}, {Marshall}, {Martino}, {Marton},
  {Mary}, {Massari}, {Matijevi{\v{c}}}, {Mazeh}, {McMillan}, {Messina},
  {Michalik}, {Millar}, {Molina}, {Molinaro}, {Moln{\'a}r}, {Montegriffo},
  {Mor}, {Morbidelli}, {Morel}, {Morris}, {Mulone}, {Muraveva}, {Musella},
  {Nelemans}, {Nicastro}, {Noval}, {O'Mullane}, {Ord{\'e}novic},
  {Ord{\'o}{\~n}ez-Blanco}, {Osborne}, {Pagani}, {Pagano}, {Pailler},
  {Palacin}, {Palaversa}, {Panahi}, {Pawlak}, {Piersimoni}, {Pineau}, {Plachy},
  {Plum}, {Poggio}, {Poujoulet}, {Pr{\v{s}}a}, {Pulone}, {Racero}, {Ragaini},
  {Rambaux}, {Ramos-Lerate}, {Regibo}, {Reyl{\'e}}, {Riclet}, {Ripepi}, {Riva},
  {Rivard}, {Rixon}, {Roegiers}, {Roelens}, {Romero-G{\'o}mez}, {Rowell},
  {Royer}, {Ruiz-Dern}, {Sadowski}, {Sagrist{\`a} Sell{\'e}s}, {Sahlmann},
  {Salgado}, {Salguero}, {Sanna}, {Santana-Ros}, {Sarasso}, {Savietto},
  {Schultheis}, {Sciacca}, {Segol}, {Segovia}, {S{\'e}gransan}, {Shih},
  {Siltala}, {Silva}, {Smart}, {Smith}, {Solano}, {Solitro}, {Sordo}, {Soria
  Nieto}, {Souchay}, {Spagna}, {Spoto}, {Stampa}, {Steele},
  {Steidelm{\"u}ller}, {Stephenson}, {Stoev}, {Suess}, {Surdej}, {Szabados},
  {Szegedi-Elek}, {Tapiador}, {Taris}, {Tauran}, {Taylor}, {Teixeira},
  {Terrett}, {Teyssand ier}, {Thuillot}, {Titarenko}, {Torra Clotet}, {Turon},
  {Ulla}, {Utrilla}, {Uzzi}, {Vaillant}, {Valentini}, {Valette}, {van Elteren},
  {Van Hemelryck}, {van Leeuwen}, {Vaschetto}, {Vecchiato}, {Veljanoski},
  {Viala}, {Vicente}, {Vogt}, {von Essen}, {Voss}, {Votruba}, {Voutsinas},
  {Walmsley}, {Weiler}, {Wertz}, {Wevers}, {Wyrzykowski}, {Yoldas},
  {{\v{Z}}erjal}, {Ziaeepour}, {Zorec}, {Zschocke}, {Zucker}, {Zurbach}, \&
  {Zwitter}}]{gaia3}
{Gaia Collaboration}, {Brown}, A.~G.~A., {Vallenari}, A., {et~al.} 2018, \aap,
  616, A1

\bibitem[{{Gaia Collaboration} {et~al.}(2016){Gaia Collaboration}, {Prusti},
  {de Bruijne}, {Brown}, {Vallenari}, {Babusiaux}, {Bailer-Jones}, {Bastian},
  {Biermann}, {Evans}, \& et~al.}]{gaia1}
{Gaia Collaboration}, {Prusti}, T., {de Bruijne}, J.~H.~J., {et~al.} 2016,
  \aap, 595, A1

\bibitem[{{Gies} {et~al.}(1998){Gies}, {Bagnuolo}, {Ferrara}, {Kaye},
  {Thaller}, {Penny}, \& {Peters}}]{gies98}
{Gies}, D.~R., {Bagnuolo}, William~G., J., {Ferrara}, E.~C., {et~al.} 1998,
  \apj, 493, 440

\bibitem[{{Hadrava}(1997)}]{korel2}
{Hadrava}, P. 1997, \aaps, 122, 581

\bibitem[{{Hadrava}(2004{\natexlab{a}})}]{fotel}
{Hadrava}, P. 2004{\natexlab{a}}, Publ. Astron. Inst. Acad. Sci. Czech Rep.,
  92, 1

\bibitem[{{Hadrava}(2004{\natexlab{b}})}]{korel}
{Hadrava}, P. 2004{\natexlab{b}}, Publ. Astron. Inst. Acad. Sci. Czech Rep.,
  92, 15

\bibitem[{{Hall} \& {Vanlandingham}(1970)}]{hall70}
{Hall}, D.~S. \& {Vanlandingham}, F.~G. 1970, \pasp, 82, 640

\bibitem[{{Harmanec}(1983)}]{hec83}
{Harmanec}, P. 1983, Hvar Observatory Bulletin, 7, 55

\bibitem[{{Harmanec}(1998)}]{hpvb}
{Harmanec}, P. 1998, \aap, 335, 173

\bibitem[{{Harmanec}(2003)}]{hec2003}
{Harmanec}, P. 2003, in New Directions for Close Binary Studies: The Royal Road
  to the Stars, ed. O.~{Demircan} \& E.~{Budding}, 221; see
  https://astro.troja.mff.cuni.cz/ftp/hec/can2002.ps

\bibitem[{{Harmanec} \& {Bo{\v{z}}i{\'c}}(2001)}]{hecboz2001}
{Harmanec}, P. \& {Bo{\v{z}}i{\'c}}, H. 2001, \aap, 369, 1140

\bibitem[{{Harmanec} {et~al.}(2002){Harmanec}, {Bo\v{z}i{\'c}}, {Percy},
  {Yang}, {Ruzdjak}, {Sudar}, {Wolf}, {Iliev}, {Huang}, {Buil}, \&
  {Eenens}}]{zarf21}
{Harmanec}, P., {Bo\v{z}i{\'c}}, H., {Percy}, J.~R., {et~al.} 2002, \aap, 387,
  580

\bibitem[{{Harmanec} {et~al.}(2000){Harmanec}, {Habuda}, {{\v{S}}tefl},
  {Hadrava}, {Kor{\v{c}}{\'a}kov{\'a}}, {Koubsk{\'y}}, {Krti{\v{c}}ka},
  {Kub{\'a}t}, {{\v{S}}koda}, {{\v{S}}lechta}, \& {Wolf}}]{zarf20}
{Harmanec}, P., {Habuda}, P., {{\v{S}}tefl}, S., {et~al.} 2000, \aap, 364, L85

\bibitem[{{Harmanec} \& {Horn}(1998)}]{hechor98}
{Harmanec}, P. \& {Horn}, J. 1998, Journal of Astronomical Data, 4, 5

\bibitem[{{Harmanec} {et~al.}(1994){Harmanec}, {Horn}, \& {Juza}}]{hhj94}
{Harmanec}, P., {Horn}, J., \& {Juza}, K. 1994, \aaps, 104, 121

\bibitem[{{Harmanec} {et~al.}(1974){Harmanec}, {Koubsk{\'y}}, \&
  {Krpata}}]{hec74}
{Harmanec}, P., {Koubsk{\'y}}, P., \& {Krpata}, J. 1974, \aap, 33, 117

\bibitem[{{Harmanec} {et~al.}(1976){Harmanec}, {Koubsk{\'y}}, {Krpata}, \&
  {{\v{Z}}d{\'a}rsk{\'y}}}]{zarf6}
{Harmanec}, P., {Koubsk{\'y}}, P., {Krpata}, J., \& {{\v{Z}}d{\'a}rsk{\'y}}, F.
  1976, Bulletin of the Astronomical Institutes of Czechoslovakia, 27, 47

\bibitem[{{Harmanec} {et~al.}(2015){Harmanec}, {Koubsk{\'y}}, {Nemravov{\'a}},
  {Royer}, {Briot}, {North}, {Lampens}, {Fr{\'e}mat}, {Yang},
  {Bo{\v{z}}i{\'c}}, {Kotkov{\'a}}, {{\v{S}}koda}, {{\v{S}}lechta},
  {Kor{\v{c}}{\'a}kov{\'a}}, {Wolf}, \& {Zasche}}]{zarf30}
{Harmanec}, P., {Koubsk{\'y}}, P., {Nemravov{\'a}}, J.~A., {et~al.} 2015, \aap,
  573, A107

\bibitem[{{Hill} {et~al.}(1976){Hill}, {Hilditch}, \&
  {Pfannenschmidt}}]{hill76}
{Hill}, G., {Hilditch}, R.~W., \& {Pfannenschmidt}, E.~L. 1976, Publications of
  the Dominion Astrophysical Observatory Victoria, 15, 1

\bibitem[{{Horn} {et~al.}(1996){Horn}, {Kub\'at}, {Harmanec}, {Koubsk\'y},
  {Hadrava}, {\v{S}imon}, {\v{S}tefl}, \& {\v{S}koda}}]{sef0}
{Horn}, J., {Kub\'at}, J., {Harmanec}, P., {et~al.} 1996, \aap, 309, 521

\bibitem[{{Jayasinghe} {et~al.}(2019){Jayasinghe}, {Stanek}, {Kochanek},
  {Shappee}, {Holoien}, {Thompson}, {Prieto}, {Dong}, {Pawlak}, {Pejcha},
  {Shields}, {Pojmanski}, {Otero}, {Hurst}, {Britt}, \& {Will}}]{asas2019}
{Jayasinghe}, T., {Stanek}, K.~Z., {Kochanek}, C.~S., {et~al.} 2019, \mnras,
  485, 961

\bibitem[{{Kervella} {et~al.}(2019){Kervella}, {Arenou}, {Mignard}, \&
  {Th{\'e}venin}}]{kervella2019}
{Kervella}, P., {Arenou}, F., {Mignard}, F., \& {Th{\'e}venin}, F. 2019, \aap,
  623, A72

\bibitem[{{Kochanek} {et~al.}(2017){Kochanek}, {Shappee}, {Stanek}, {Holoien},
  {Thompson}, {Prieto}, {Dong}, {Shields}, {Will}, {Britt}, {Perzanowski}, \&
  {Pojma{\'n}ski}}]{asas2017}
{Kochanek}, C.~S., {Shappee}, B.~J., {Stanek}, K.~Z., {et~al.} 2017, \pasp,
  129, 104502

\bibitem[{{Koen} \& {Eyer}(2002)}]{koen2002}
{Koen}, C. \& {Eyer}, L. 2002, \mnras, 331, 45

\bibitem[{{Krpata}(2008)}]{spefo3}
{Krpata}, J. 2008, http://astro.troja.mff.cuni.cz/ftp/hec/SPEFO/

\bibitem[{{Labadie-Bartz} {et~al.}(2017){Labadie-Bartz}, {Pepper}, {McSwain},
  {Bjorkman}, {Bjorkman}, {Lund}, {Rodriguez}, {Stassun}, {Stevens}, {James},
  {Kuhn}, {Siverd}, \& {Beatty}}]{bartz2017}
{Labadie-Bartz}, J., {Pepper}, J., {McSwain}, M.~V., {et~al.} 2017, \aj, 153,
  252

\bibitem[{{Lanz} \& {Hubeny}(2007)}]{bgrid}
{Lanz}, T. \& {Hubeny}, I. 2007, \apjs, 169, 83

\bibitem[{{Lightkurve Collaboration} {et~al.}(2018){Lightkurve Collaboration},
  {Cardoso}, {Hedges}, {Gully-Santiago}, {Saunders}, {Cody}, {Barclay}, {Hall},
  {Sagear}, {Turtelboom}, {Zhang}, {Tzanidakis}, {Mighell}, {Coughlin}, {Bell},
  {Berta-Thompson}, {Williams}, {Dotson}, \& {Barentsen}}]{Lightkurve}
{Lightkurve Collaboration}, {Cardoso}, J. V. d. M.~a., {Hedges}, C., {et~al.}
  2018, {Lightkurve: Kepler and TESS time series analysis in Python}

\bibitem[{{Neiner} {et~al.}(2011){Neiner}, {de Batz}, {Cochard}, {Floquet},
  {Mekkas}, \& {Desnoux}}]{neiner2011}
{Neiner}, C., {de Batz}, B., {Cochard}, F., {et~al.} 2011, \aj, 142, 149

\bibitem[{{Nemravov{\'a}} {et~al.}(2016){Nemravov{\'a}}, {Harmanec},
  {Bro{\v{z}}}, {Vokrouhlick{\'y}}, {Mourard}, {Hummel}, {Cameron}, {Matthews},
  {Bolton}, {Bo{\v{z}}i{\'c}}, {Chini}, {Dembsky}, {Engle}, {Farrington},
  {Grunhut}, {Guenther}, {Guinan}, {Kor{\v{c}}{\'a}kov{\'a}}, {Koubsk{\'y}},
  {K{\v{r}}{\'\i}{\v{c}}ek}, {Kuschnig}, {Mayer}, {McCook}, {Moffat},
  {Nardetto}, {Pr{\v{s}}a}, {Ribeiro}, {Rowe}, {Rucinski}, {{\v{S}}koda},
  {{\v{S}}lechta}, {Tallon-Bosc}, {Votruba}, {Weiss}, {Wolf}, {Zasche}, \&
  {Zavala}}]{jn2016}
{Nemravov{\'a}}, J.~A., {Harmanec}, P., {Bro{\v{z}}}, M., {et~al.} 2016, \aap,
  594, A55

\bibitem[{{Palacios} {et~al.}(2010){Palacios}, {Gebran}, {Josselin}, {Martins},
  {Plez}, {Belmas}, \& {L{\`e}bre}}]{pala2010}
{Palacios}, A., {Gebran}, M., {Josselin}, E., {et~al.} 2010, \aap, 516, A13

\bibitem[{{Pepper} {et~al.}(2007){Pepper}, {Pogge}, {DePoy}, {Marshall},
  {Stanek}, {Stutz}, {Poindexter}, {Siverd}, {O'Brien}, {Trueblood}, \&
  {Trueblood}}]{Pepper2007}
{Pepper}, J., {Pogge}, R.~W., {DePoy}, D.~L., {et~al.} 2007, \pasp, 119, 923

\bibitem[{{Perryman} \& {ESA}(1997)}]{esa97}
{Perryman}, M.~A.~C. \& {ESA}. 1997, {The HIPPARCOS and TYCHO catalogues}
  (Astrometric and photometric star catalogues derived from the ESA Hipparcos
  Space Astrometry Mission, Publisher: Noordwijk, Netherlands: ESA Publications
  Division, 1997, Series: ESA SP Series 1200)

\bibitem[{{Plaskett} \& {Pearce}(1931)}]{plaskett}
{Plaskett}, J.~S. \& {Pearce}, J.~A. 1931, Publications of the Dominion
  Astrophysical Observatory Victoria, 5, 1

\bibitem[{{Pojmanski}(2002)}]{pojm2002}
{Pojmanski}, G. 2002, Acta Astronomica, 52, 397

\bibitem[{{Pr{\v{s}}a} {et~al.}(2016){Pr{\v{s}}a}, {Harmanec}, {Torres},
  {Mamajek}, {Asplund}, {Capitaine}, {Christensen-Dalsgaard}, {Depagne},
  {Haberreiter}, {Hekker}, {Hilton}, {Kopp}, {Kostov}, {Kurtz}, {Laskar},
  {Mason}, {Milone}, {Montgomery}, {Richards}, {Schmutz}, {Schou}, \&
  {Stewart}}]{units2016}
{Pr{\v{s}}a}, A., {Harmanec}, P., {Torres}, G., {et~al.} 2016, \aj, 152, 41

\bibitem[{{Ricker} {et~al.}(2016){Ricker}, {Vanderspek}, {Winn}, {Seager},
  {Berta-Thompson}, {Levine}, {Villasenor}, {Latham}, {Charbonneau}, {Holman},
  {Johnson}, {Sasselov}, {Szentgyorgyi}, {Torres}, {Bakos}, {Brown},
  {Christensen-Dalsgaard}, {Kjeldsen}, {Clampin}, {Rinehart}, {Deming}, {Doty},
  {Dunham}, {Ida}, {Kawai}, {Sato}, {Jenkins}, {Lissauer}, {Jernigan},
  {Kaltenegger}, {Laughlin}, {Lin}, {McCullough}, {Narita}, {Pepper},
  {Stassun}, \& {Udry}}]{Ricker2016}
{Ricker}, G.~R., {Vanderspek}, R., {Winn}, J., {et~al.} 2016, Society of
  Photo-Optical Instrumentation Engineers (SPIE) Conference Series, Vol. 9904,
  {The Transiting Exoplanet Survey Satellite}, 99042B

\bibitem[{{Rivinius} {et~al.}(2013){Rivinius}, {Carciofi}, \&
  {Martayan}}]{rivi2013}
{Rivinius}, T., {Carciofi}, A.~C., \& {Martayan}, C. 2013, \aapr, 21, 69

\bibitem[{{Schaller} {et~al.}(1992){Schaller}, {Schaerer}, {Meynet}, \&
  {Maeder}}]{schaller92}
{Schaller}, G., {Schaerer}, D., {Meynet}, G., \& {Maeder}, A. 1992, \aaps, 96,
  269

\bibitem[{{Shappee} {et~al.}(2014){Shappee}, {Prieto}, {Grupe}, {Kochanek},
  {Stanek}, {De Rosa}, {Mathur}, {Zu}, {Peterson}, {Pogge}, {Komossa}, {Im},
  {Jencson}, {Holoien}, {Basu}, {Beacom}, {Szczygie{\l}}, {Brimacombe},
  {Adams}, {Campillay}, {Choi}, {Contreras}, {Dietrich}, {Dubberley},
  {Elphick}, {Foale}, {Giustini}, {Gonzalez}, {Hawkins}, {Howell}, {Hsiao},
  {Koss}, {Leighly}, {Morrell}, {Mudd}, {Mullins}, {Nugent}, {Parrent},
  {Phillips}, {Pojmanski}, {Rosing}, {Ross}, {Sand}, {Terndrup}, {Valenti},
  {Walker}, \& {Yoon}}]{asas2014}
{Shappee}, B.~J., {Prieto}, J.~L., {Grupe}, D., {et~al.} 2014, \apj, 788, 48

\bibitem[{{Sigut} \& {Patel}(2013)}]{sigut2013}
{Sigut}, T.~A.~A. \& {Patel}, P. 2013, \apj, 765, 41

\bibitem[{{Skiff}(1998)}]{skiff98}
{Skiff}, B.~A. 1998, \skytel, 95, 65

\bibitem[{{\v{S}koda}(1996)}]{spefo}
{\v{S}koda}, P. 1996, in ASP Conf. Ser. 101: Astronomical Data Analysis
  Software and Systems V, 187--189

\bibitem[{{Sterne}(1941)}]{sterne41}
{Sterne}, T.~E. 1941, Proceedings of the National Academy of Science, 27, 168

\bibitem[{{van Leeuwen}(2007{\natexlab{a}})}]{leeuw2007b}
{van Leeuwen}, F. 2007{\natexlab{a}}, in Astrophysics and Space Science
  Library, Vol. 350, Astrophysics and Space Science Library, ed. {F.~van
  Leeuwen}

\bibitem[{{van Leeuwen}(2007{\natexlab{b}})}]{leeuw2007a}
{van Leeuwen}, F. 2007{\natexlab{b}}, \aap, 474, 653

\bibitem[{{Vennes} {et~al.}(2011){Vennes}, {Kawka}, {Joni{\'c}},
  {Pirkovi{\'c}}, {Iliev}, {Kub{\'a}t}, {{\v{S}}lechta}, {N{\'e}meth}, \&
  {Kraus}}]{vennes}
{Vennes}, S., {Kawka}, A., {Joni{\'c}}, S., {et~al.} 2011, \mnras, 413, 2760

\bibitem[{{Walker} {et~al.}(1979){Walker}, {Yang}, \& {Fahlman}}]{wyf79}
{Walker}, G.~A.~H., {Yang}, S., \& {Fahlman}, G.~G. 1979, \apj, 233, 199

\bibitem[{{Wang} {et~al.}(2018){Wang}, {Gies}, \& {Peters}}]{wang2018}
{Wang}, L., {Gies}, D.~R., \& {Peters}, G.~J. 2018, \apj, 853, 156

\bibitem[{{Yamashita} {et~al.}(1977){Yamashita}, {Ichimura}, {Nakagiri},
  {Norimoto}, {Maehara}, \& {Miyajima}}]{yama77}
{Yamashita}, Y., {Ichimura}, K., {Nakagiri}, M., {et~al.} 1977, \pasj, 29, 527

\end{thebibliography}


\begin{appendix}
\section{Details on the photometric data reductions and homogenisation}\label{apa}
\begin{table*}[!tbp]
\caption[]{ Accurate Hvar all-sky mean \ubv\ values for all comparison stars used. These were added to the magnitude differences var.$-$comp. and check$-$comp.
for data from all stations.}\label{comp}
\begin{center}
\begin{tabular}{rcrcccrr}
\hline\hline\noalign{\smallskip}
Star& HD & No. of& $V$  &  $B$ &  $U$ &$(B-V)$& $(U-B)$ \\
    &    &   obs.&(mag.)&(mag.)&(mag.)&(mag.) &  (mag.) \\
\noalign{\smallskip}\hline\noalign{\smallskip}
    5 Vul&182919&  3& 5.65\p0.010& 5.66\p0.011&5.62\p0.012& 0.01& -0.04\\
    9 Vul&184606& 30& 5.010\p0.009& 4.906\p0.011&4.499\p0.012&-0.104&-0.407\\
 V395 Vul&187811& 24& 4.928\p0.015& 4.759\p0.017&4.104\p0.020&-0.170&-0.654\\
   13 Vul&188260& 38& 4.584\p0.008& 4.536\p0.010&4.404\p0.012&-0.048&-0.132\\
\noalign{\smallskip}\hline\noalign{\smallskip}
\end{tabular}
\end{center}
\end{table*}

Since we used photometry from several sources and photometric systems,
both all-sky and differential, relative to several different
comparison stars, we attempted to arrive at some homogenisation and
standardisation. We derived improved all-sky values for all comparison stars used,
employing carefully standardised \ubv\ observations secured
at Hvar specifically for this purpose. The adopted values are collected in Table~\ref{comp}
together with the number of all-sky observations and the rms errors of
one observation. These were added to the respective magnitude differences
to obtain directly comparable standard \ubv\ magnitudes for all stations.

Below, we provide some details of the individual data sets and their reductions.
\begin{itemize}
\item {\sl Station 01 -- Hvar:} \ \
 These differential observations were secured by PH and HB, and the more
recent ones also by students A.~Opli\v{s}tilov\'a and
K.~Vitovsk\'y, relative to either 9~Vul or 13~Vul.
They were carefully transformed to the standard $UBV$ system via non-linear
transformation formul\ae\ using the {\tt HEC22} reduction program; see
\citet{hhj94} and \citet{hechor98} for the observational strategy and
data reduction. \footnote{The whole program suite with a detailed manual,
examples of data, auxiliary data files, and results is available at
\url{http://astro.troja.mff.cuni.cz/ftp/hec/PHOT}\,.}
All observations were reduced with the latest
{\tt HEC22 rel.18} program, which allows the time variation of
linear extinction coefficients to be modelled in the course of observing nights. The whole archive of Hvar photometry since 1972 will be published
elsewhere after a thorough revision (Bo\v{z}i\'c et al.; in prep.).
\item {\sl Station 10 -- Kitt Peak (KPNO):}
\ \  These are all-sky \ubv\ observations published only as the mean values
\centerline{$V$=6\m31, \bv=$-0$\m10, and \ub=$-0$\m53}
\noindent and without times of observations. These were secured
at some point between RJD~37700 and 40900.
We have to accept these published values as they are because none
of the comparisons observed at Hvar have been observed (the only
exception being the variable V395~Vul).
\item {\sl Station 13 -- Mt. Kobau:} \ \ These differential observations were
obtained in the DAO (33), (44), and (55) medium-band system. We transformed
them into the Johnson \ubv\ system using the transformation formul\ae\
devised by \citet{hecboz2001}. Unfortunately, another Be star V395~Vul was
used as the comparison star for these observations. However, it seems that
the secular variations of V395~Vul are not large. Its photometry by
\citet{craw71} gives $V = 4$\m94, \bv$ = -0$\m16, and \ub $= -0$\m66, which are
values quite close to the Hvar photometry; see Table~\ref{comp}. Nevertheless,
the Hipparcos \hp\ photometry shows variability of V395~Vul on about 0\m1 level
\citep{esa97}.
\item {\sl Station 42 -- Dyer:} \ \ These differential observations relative to 5~Vul
were obtained around RJD~40500 and published as the mean values
\centerline{$V$=6\m39, \bv=$-0$\m12, and \ub=$-0$\m52.}
\noindent They were corrected for extinction and transformed to the standard Johnson \ubv\ system by the original authors.
\item {\sl Station 61 -- Hipparcos \hp:} \ \
These space observations were extracted from the ESA archive \citep{esa97} and
transformed to the standard Johnson $V$ magnitude using the transformation formula by
\citet{hpvb}. Only observations with error flags 0 and 1 were used.
\item {\sl Station 93 -- ASAS3 V photometry:} \ \ We extracted these
all-sky observations from the ASAS3 public archive \citep{pojm2002}
using the data for diaphragm~1, which had on average the lowest rms errors.
We omitted all observations of grade D and observations with rms errors
larger than 0\m04. We also omitted a strongly deviating observation at
HJD~2452662.6863.
\item {\sl Station 110 -- TESS satellite} \ \ These all-sky broad-band
observations are publicly available via the Mikulski Archive for Space
Telescopes (MAST).\footnote{
\url{http://archive.stsci.edu/tess/all_products.html}} The initial
reductions of these were carried out by one of the present authors (J.L-B.). We note that \va was observed in
TESS sector 14, and the light curve was extracted from the full-frame
images at a cadence of 30 minutes using the \textsc{lightkurve} package.

\item {\sl Station 112 -- Kutztown:} \ \ These all-sky \ubv\ data are
published as the mean values
\centerline{$V$=6\m32, \bv=$-0$\m09, and \ub=$-0$\m56}
\noindent and were secured during 17 nights during the approximate epoch RJD~42700.
\item {\sl Station 113 -- Mitaka:} \ \  These 100 all-sky observations secured
during 18 nights between Oct 22 and Dec 26, 1976 (RJD $\sim 43100$) are
again published as the mean values only:
\centerline{$V$=6\m33, \bv=$-0$\m07, and \ub=$-0$\m54.}
\item {\sl Station 114 -- ASAS-SN V photometry:} \ \ We extracted these all-sky
observations from the ASAS-SN archive. They are all from the bc camera in Hawaii and their scatter is not negligible. We removed all observations with an rms error higher
than 0\m009 and used only observations within the magnitude range from 6\m65
to 6\m80. Finally, comparing the robust mean value from the Hvar calibrated
$V$ photometry and from the cleaned ASAS-SN observations, we derived a correction
of $-0$\m380 to bring the ASAS-SN data to the mean Hvar $V$ magnitude.
\item {\sl Station 115 -- KELT (Kilodegree Extremely Little Telescope)}
at Winer Observatory, USA; broad-band $R$ photometry: We adopted
these observations from the paper of \citet{bartz2017} and decreased
all observed values for 0\m856 to bring them into the range
of $V$ photometry for our analyses.
\end{itemize}

\section{Details of the spectral data reduction and measurements}\label{apb}
The initial reduction of all OND and DAO spectra (bias subtraction,
flat-fielding, creation of 1D spectra, and wavelength calibration) was
carried out in {\tt IRAF}. Optimal extraction was used.
Normalisation and RV measurements of \ha absorption were carried out with
the program \respefoe. The RV measurements are based on the comparison of direct and flipped line profiles in normalised spectra.
Additionally, we also measured a selection of unblended water vapour
lines and their robust mean RVs were used to fine corrections of the
RV zero point of each spectrogram.

\begin{table*}
\caption[]{Individual RVs (in \ks) of the \ha absorption measured
on new DAO, OND, and BeSS spectra with \respefoe.}
\label{newrv}
\begin{center}
\begin{tabular}{crrcrrcccccl}
\hline\hline\noalign{\smallskip}
RJD&RV&Observatory&RJD&RV&Observatory\\
\noalign{\smallskip}\hline\noalign{\smallskip}
54276.8204&  -13.17& DAO& 55373.5085&   -8.85& OND\\
54339.7281&   -7.15& DAO& 55377.3711&   -7.60& OND\\
54340.7416&   -6.57& DAO& 55377.3836&   -6.90& OND\\
54367.7330&  -23.33& DAO& 55429.3155&  -18.60& OND\\
54443.6185&  -26.81& DAO& 55429.3230&  -18.68& OND\\
54244.4463&  -22.06& OND& 55430.3259&  -18.92& OND\\
54286.4843&  -14.41& OND& 55430.3334&  -19.28& OND\\
54288.3624&  -17.14& OND& 55431.3336&  -17.05& OND\\
54288.5549&  -16.67& OND& 55431.3413&  -19.66& OND\\
54289.5548&  -17.64& OND& 55461.3135&  -14.06& OND\\
54295.4775&  -20.14& OND& 55462.3026&  -17.46& OND\\
54295.5953&  -20.54& OND& 55463.2886&  -17.46& OND\\
54296.3730&  -21.08& OND& 55464.2872&  -15.35& OND\\
54296.5439&  -22.12& OND& 55470.2794&  -20.34& OND\\
54296.5632&  -21.70& OND& 55483.2938&  -27.65& OND\\
54296.5809&  -21.54& OND& 55625.7053&  -25.15& OND\\
54297.4056&  -21.85& OND& 55648.5782&   -8.82& OND\\
54297.4426&  -22.57& OND& 55648.5925&  -10.28& OND\\
54297.5462&  -22.20& OND& 55650.6278&   -8.71& OND\\
54297.5813&  -22.60& OND& 55675.5853&  -17.01& OND\\
54304.4063&  -24.42& OND& 55777.3572&  -20.56& OND\\
54304.5040&  -25.79& OND& 55834.3867&  -27.21& OND\\
54307.3585&  -26.78& OND& 55878.1764&  -14.34& OND\\
54308.3633&  -26.39& OND& 55878.2884&  -11.57& OND\\
54309.5100&  -24.20& OND& 55879.2138&  -14.85& OND\\
54317.5246&  -18.69& OND& 55879.3118&  -14.66& OND\\
54318.5311&  -17.62& OND& 55906.1654&  -26.60& OND\\
54319.3650&  -17.22& OND& 55906.2346&  -26.45& OND\\
54325.4170&  -12.39& OND& 58640.3665&   -6.74& OND\\
54387.2461&  -20.54& OND& 58640.3993&  -11.07& OND\\
54387.3810&  -18.83& OND& 58640.5103&  -10.34& OND\\
54389.2323&  -16.46& OND& 58648.3710&   -6.09& OND\\
54389.2493&  -17.62& OND& 58649.3894&   -5.20& OND\\
54389.3731&  -17.05& OND& 58651.4545&   -7.56& OND\\
54390.2819&  -14.70& OND& 58654.4355&   -9.83& OND\\
54649.5466&  -26.16& OND& 58681.3902&  -24.72& OND\\
54706.5216&  -17.80& OND& 58681.4217&  -29.31& OND\\
54753.2694&   -6.13& OND& 58682.3705&  -20.78& OND\\
54753.2881&   -5.61& OND& 58689.4461&  -18.51& OND\\
54934.5948&  -25.90& OND& 58689.4575&  -21.33& OND\\
55043.4394&  -12.49& OND& 58690.3439&  -25.72& OND\\
55043.4541&  -13.49& OND& 58690.4058&  -22.31& OND\\
55044.3682&  -14.48& OND& 58691.3792&  -24.45& OND\\
55045.5963&  -15.50& OND& 58714.3611&  -11.60& OND\\
55074.3366&  -25.84& OND& 56480.4756&  -15.19& BeSS\\
55074.4756&  -25.65& OND& 56817.5138&  -16.89& BeSS\\
55096.3567&   -7.21& OND& 57224.5244&  -22.53& BeSS\\
55098.2608&   -6.49& OND& 57579.5281&  -27.70& BeSS\\
55098.4263&   -6.58& OND& 57676.5577&  -11.72& BeSS\\
55098.4422&   -5.90& OND& 58729.6276&  -18.33& BeSS\\
55114.2697&  -14.76& OND& 56508.3994&  -19.46& BeSS\\
55114.3802&  -13.68& OND& 58310.4416&   -9.25& BeSS\\
55310.6018&   -6.06& OND& 58338.3540&  -19.00& BeSS\\
55310.6162&   -8.16& OND& 58663.4813&  -17.09& BeSS\\
55311.6199&   -7.29& OND& 57205.3757&  -14.29& BeSS\\
55312.6144&   -8.33& OND& 57970.3965&  -15.72& BeSS\\
55314.5922&   -8.81& OND& 57207.5716&  -20.73& BeSS\\
55314.6030&   -7.15& OND& 58258.4909&  -22.72& BeSS\\
\noalign{\smallskip}\hline\noalign{\smallskip}
\end{tabular}
\end{center}
\end{table*}

\begin{table*}[!tbp]
\caption[]{Individual RVs of the primary from old DAO spectra
\citep{plaskett}. We derived RJDs and averaged RVs in cases where the same
plate was measured several times. The RVs are mean values based on
measurements of 5 to 12 spectral lines of \ion{h}{i} and \ion{He}{i}
in the blue spectral region.}
\label{oldrv}
\begin{center}
\begin{tabular}{ccrccrrcccccl}
\hline\hline\noalign{\smallskip}
RJD&mean RV&No. of&RJD&mean RV&No. of\\
   & (\ks) &lines &   &(\ks)  &lines\\
\noalign{\smallskip}\hline\noalign{\smallskip}
22185.7940&-46.5& 5-7&23992.7564&-43.7&  9 \\
22249.5973&-46.0&7-10&24000.7913&-41.0&  9 \\
22537.8743&-42.4& 10 &24359.8474&-41.4& 12  \\
22549.7940&-43.2& 8  \\
\noalign{\smallskip}\hline\noalign{\smallskip}
\end{tabular}
\end{center}
\end{table*}

\section{The Java reduction program reSPEFO}\label{apc}
\respefo is a modern refresh of the original program \spefoe, a~simple yet powerful program for 1D spectra analysis written in Turbo Pascal \citep{sef0,spefo,spefo3}.

The new version of reSPEFO is written in Java using contemporary tools with an improved user interface and functionality.
The program is distributed as a runnable \verb|.jar| archive.
It has minimal system requirements and runs on most Linux, Windows, and MacOS systems.

\begin{figure*}[!tbp]
  \centering
  \subfloat[Original spectrum]{\includegraphics[width=0.6\textwidth]{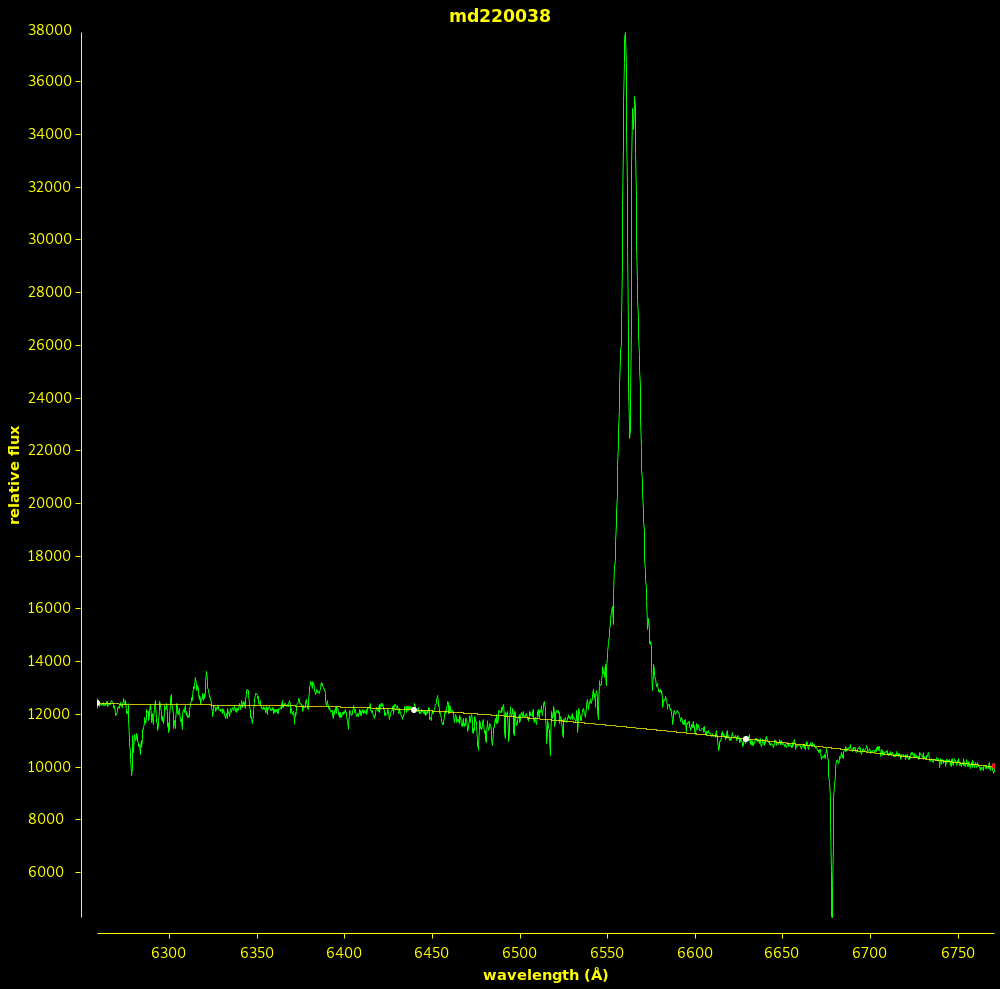}\label{rec1}}
  \hfill
  \subfloat[normalised spectrum]{\includegraphics[width=0.6\textwidth]{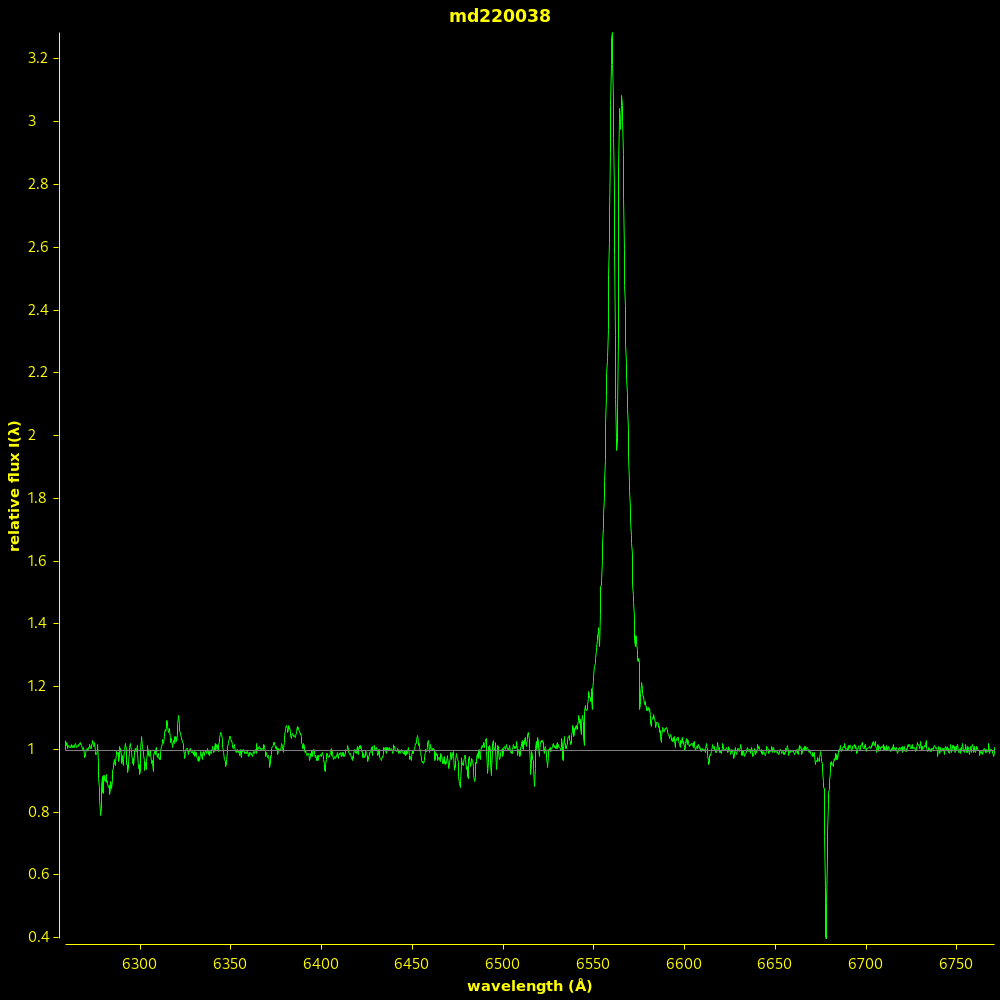}\label{rec2}}
\caption{Interactive spectra normalisation. A spectrum of V1294~Aql,
a~Be star with strong \ha emission, is shown here.}
  \label{rec}
\end{figure*}

\begin{figure*}[!tbp]
  \centering
  \includegraphics[width=\textwidth]{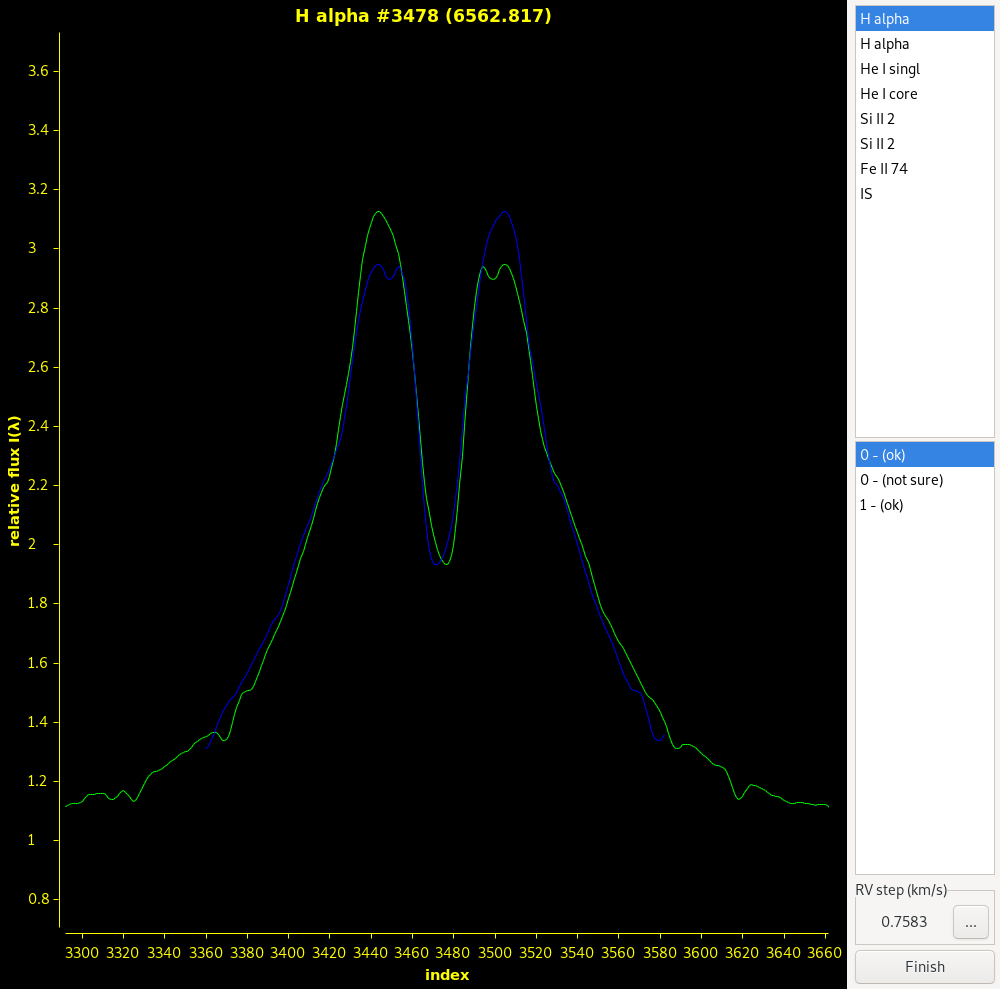}
  \caption{Interactive RV measurements: RV of the central absorption core
is measured in this example. It is seen that when one would set on the
outer emission-line wings, a different RV would be obtained. Measuring points are defined in a separate file using an identification and a  laboratory wavelength as shown in the list in the top-right corner. The user can make multiple measurements on a single point supplemented with an optional comment as shown in the list in the bottom-right corner. Once again, a spectrum of V1294~Aql was used here.}
  \label{rv}
\end{figure*}

\begin{figure*}[!tbp]
  \centering
  \includegraphics[width=\textwidth]{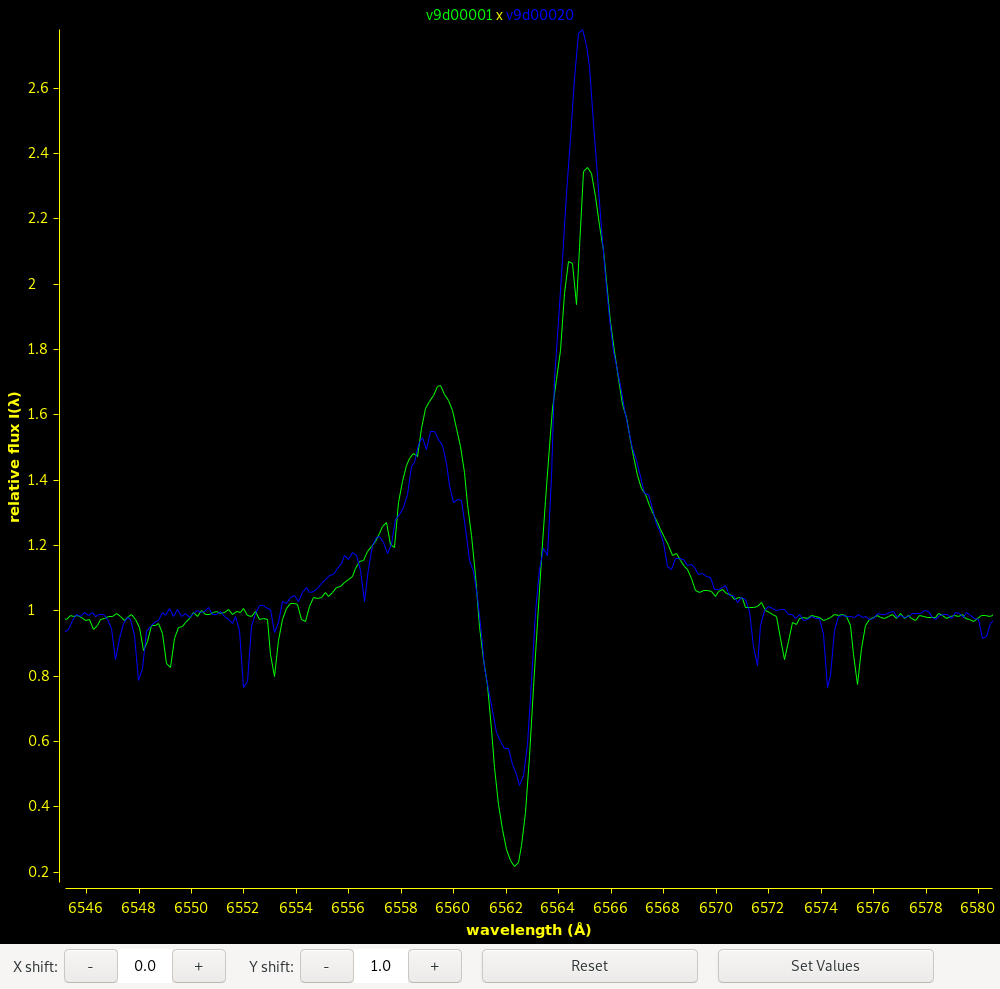}
  \caption{Comparison of two DAO spectra of \ve. Mutual shifts of the profiles can be carried out interactively and the difference in RV is
  displayed. It is also possible to change the relative strength of the second spectrum (shown in blue).}
  \label{compare}
\end{figure*}

It consists of a suite of tools for operations with linear spectra.
The current version provides these functions:
\begin{enumerate}
\item Creation of a list of spectra available in a given directory,
which contains running numbers of the spectra in an~increasing order
in time, their original names, and HJDs;
\item Converting the names of all spectra into a~format compatible with
the DOS convention used in the original \spefoe, i.e. three characters representing the name of the set of spectra, a five-digit running number, and the extension describing the type of spectra;
\item Import of spectra recorded in the original internal format of
\spefoe;
\item Interactive rectification of spectra;
\item Interactive cleaning of cosmic ray hits and other flaws;
\item Comparison of two selected spectra with a possibility
to move the second spectrum in RV and proportionally reduce its flux;
\item Interactive RV measurements based on a~comparison of direct and
flipped line-profile images. Individual lines of groups of lines can be
put into different user-defined groups;
\item Export of all RV measurements into a table containing HJDs
and RVs of individual groups of lines chosen by the user (if telluric
lines were also measured, a file with a fine correction of the RV zero point is  also created); and
\item Export of the spectra into either FITS or ASCII format.
\end{enumerate}
Screenshots illustrating some of these features are provided in
Figs.~\ref{rec}, \ref{rv}, and \ref{compare}.

The application \respefo is available for free under the Eclipse Public License 2.0 at \url{https://astro.troja.mff.cuni.cz/projects/respefo}, where
a detailed User Guide is  also provided.

\end{appendix}
\end{document}